\documentclass[preprint,journal]{vgtc}            %

\onlineid{1469}

\vgtccategory{Systems \& Rendering}

\vgtcpapertype{System}

\newcommand\SoftwareName{\mbox{ManiVault}} %

\title{\SoftwareName: A Flexible and Extensible \break Visual Analytics Framework for High-Dimensional Data}

\makeatletter
\let\textTitle\@title
\makeatother

\author{%
Alexander Vieth$^{\ast 1}$\orcidlink{0000-0002-5809-4316}, %
Thomas Kroes$^{\ast 2}$\orcidlink{0000-0002-0658-2203}, %
Julian Thijssen$^{\ast 2}$\orcidlink{0009-0007-7585-6451}, %
Baldur van Lew$^{2}$\orcidlink{0000-0003-0628-1264},  %
Jeroen Eggermont$^{2}$\orcidlink{0000-0003-0361-9469}, \\ %
Soumyadeep Basu$^{2}$\orcidlink{0000-0002-8918-9203}, %
Elmar Eisemann$^{1}$\orcidlink{0000-0003-4153-065X}, %
Anna Vilanova$^{3}$\orcidlink{0000-0002-1034-737X}, %
Thomas Höllt$^{\bullet 1}$\orcidlink{0000-0001-8125-1650}, %
Boudewijn Lelieveldt$^{\bullet 1,2}$\orcidlink{0000-0001-8269-7603}
}

\authorfooter{
\item[$^{\ast \bullet}$] These authors contributed equally.
\item[$^{1}$] TU Delft, E-mail: \{A.Vieth~$|$~E.Eisemann~$|$~T.Hollt-1\} @tudelft.nl.
\item[$^{2}$] Leiden University Medical Center, E-mail: \{J.G.L.Thijssen~$|$~T.Kroes~$|$ J.Eggermont~$|$~B.van\_Lew~$|$~S.Basu~$|$~B.P.F.Lelieveldt\} @lumc.nl.
\item[$^{3}$] TU Eindhoven, E-mail: A.Vilanova@tue.nl.
}

\makeatletter
\def\convertto#1#2{\strip@pt\dimexpr #2*65536/\number\dimexpr 1#1}
\makeatother

\newcommand{\getHeight}[1]{%
	\newsavebox{\mytext}
	\newlength{\heightmytext}
	\newlength{\depthmytext}
	\newlength{\totaltextheight}
	\newlength{\abswidth}%
	\setlength{\abswidth}{\textwidth}%
	\addtolength{\abswidth}{-.66in}%
	\savebox{\mytext}{%
		\parbox{\abswidth}{
			#1
		}%
	}%
\settoheight{\heightmytext}{\usebox{\mytext}}
\settodepth{\depthmytext}{\usebox{\mytext}}
\settototalheight{\totaltextheight}{\usebox{\mytext}}
\noindent\par
Text height is \the\totaltextheight, which are \convertto{in}{ \the\totaltextheight} in or \convertto{cm}{ \the\totaltextheight} cm. Max is 3 in (7.62 cm).
}%

\abstract{%
Exploration and analysis of high-dimensional data are important tasks in many fields that produce large and complex data, like the financial sector, systems biology, or cultural heritage.
Tailor-made visual analytics software is developed for each specific application, limiting their applicability in other fields.
However, as diverse as these fields are, their characteristics and requirements for data analysis are conceptually similar.
Many applications share abstract tasks and data types and are often constructed with similar building blocks.
Developing such applications, even when based mostly on existing building blocks, requires significant engineering efforts.
We developed \SoftwareName, a flexible and extensible open-source visual analytics framework for analyzing high-dimensional data. 
The primary objective of \SoftwareName\ is to facilitate rapid prototyping of visual analytics workflows for visualization software developers and practitioners alike.
\SoftwareName\ is built using a plugin-based architecture that offers easy extensibility.
While our architecture deliberately keeps plugins self-contained, to guarantee maximum flexibility and re-usability, we have designed and implemented a messaging API for tight integration and linking of modules to support common visual analytics design patterns.
We provide several visualization and analytics plugins, and \SoftwareName's API makes the integration of new plugins easy for developers. 
\SoftwareName\ facilitates the distribution of visualization and analysis pipelines and results for practitioners through saving and reproducing complete application states.
As such, \SoftwareName\ can be used as a communication tool among researchers to discuss workflows and results.  \\
A copy of this paper and all supplemental material is available at \href{https://osf.io/9k6jw/}{osf.io/9k6jw}, and source code at \href{https://github.com/ManiVaultStudio}{github.com/ManiVaultStudio}.
}

\keywords{%
High-dimensional data, %
Visual analytics, %
Visualization framework, %
Progressive analytics, %
Prototyping system. %
}

\teaser{%

\centering
\pdftooltip{%
\includegraphics[width=\linewidth]{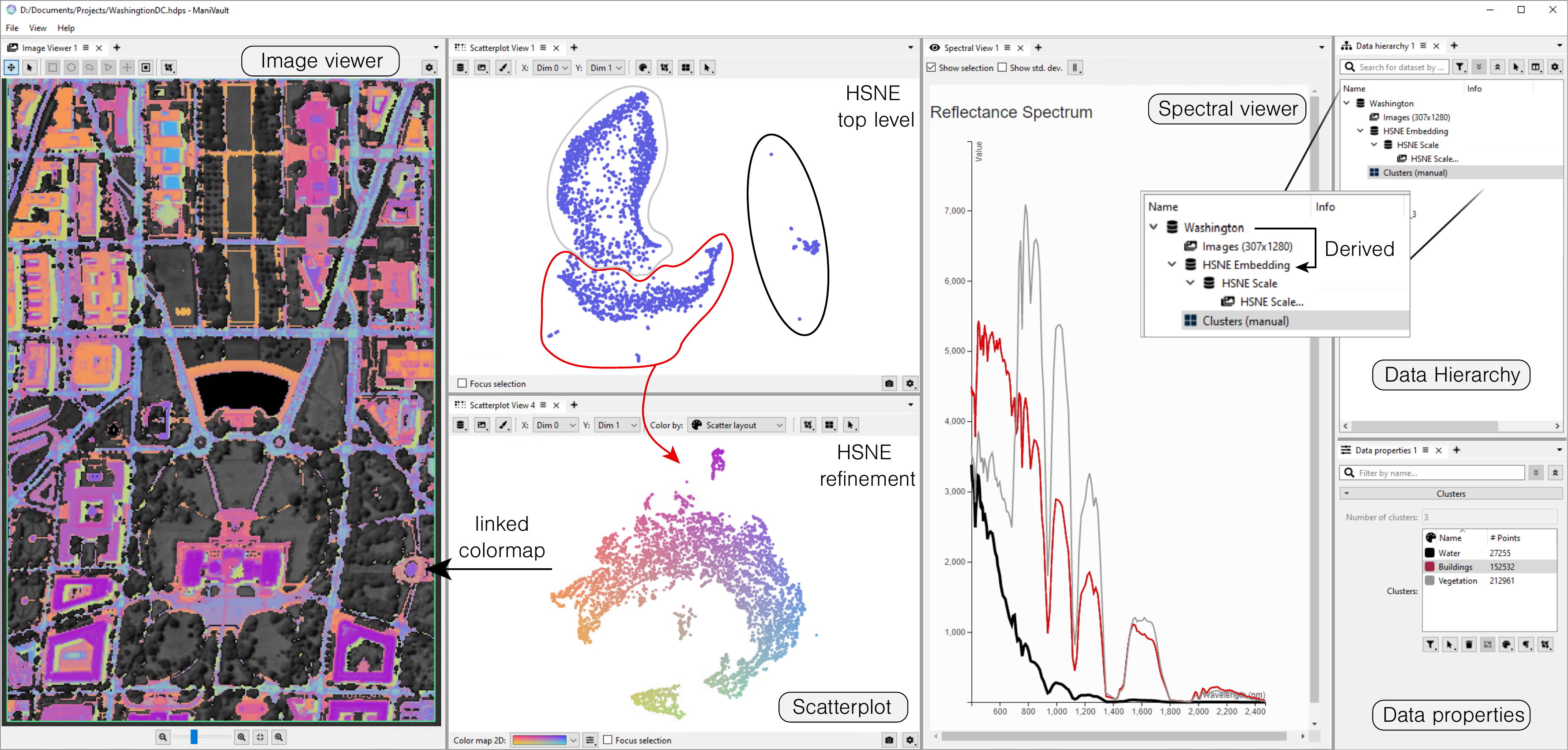}
\vspace{-6mm}
}{Screenshot of ManiVault with multiple view plugins showing various primary and derived data sets used for the exploration of a hyperspectral image.}
\caption{%
\textbf{Example screenshot of \SoftwareName{}} used for the exploration of a hyperspectral imaging data set.%
}
\label{fig:teaser}

}

\graphicspath{{figures/}{tables/}{./}} %

\usepackage{mathptmx}                  %
\usepackage[svgnames, table]{xcolor}   %

\usepackage{comment}            %
\usepackage{multirow}           %
\usepackage{booktabs}           %
\usepackage[referable]{threeparttablex}
\usepackage{amsmath}
\usepackage{pdfcomment}         %
\usepackage{listings}

\definecolor{codegreen}{rgb}{0.04,0.6,0.02}
\definecolor{codered}{rgb}{1,0.18,0}
\definecolor{codepurple}{rgb}{0.58,0,0.82}
\definecolor{codeblue}{rgb}{0.23,0.08,1}
\definecolor{codegray}{rgb}{0.38,0.38,0.38}
\definecolor{codebackground}{rgb}{0.98,0.98,0.98}

\lstdefinestyle{CppStyle}{%
  language=[ISO]C++,
  frame=tb, 
  numbers=none,
  commentstyle=\color{codegreen},
  keywordstyle=\color{codepurple},
  directivestyle=\color{black},
  numberstyle=\tiny\color{codegray},
  stringstyle=\color{codered},
  backgroundcolor=\color{codebackground},   
  emph={int,char,double,float,unsigned},
  emphstyle=\color{codepurple},
  basicstyle=\ttfamily\footnotesize, %
  breaklines=true,
  keepspaces=true,
  escapeinside={(*@}{@*)},
}
\usepackage{siunitx}            %
\usepackage[fixed]{fontawesome5}%
\usepackage{stfloats}
\usepackage{threeparttable}
\usepackage{placeins}
\usepackage{orcidlink}

\input{utils/autorefAdditions}

\newcommand\checkMaxPages[1]{
\ifnum\thepage>#1
    \noindent
    \colorbox{Orange}{
    \parbox{\columnwidth}{
    MAX #1 PAGES ALLOWED FOR CONTENT \\
    (excluding acknowledgments, references and figure credits.) \\
    This is page \thepage. That's too much.
    }}
\fi
}

\newcommand*{\reqref}[1]{%
  \hyperref[{#1}]{%
    Requirement~\ref*{#1}%
  }%
}

\newcommand*{\usrref}[1]{%
  \hyperref[{#1}]{%
    User~\ref*{#1}%
  }%
}

\usepackage{tikz}
\usepackage{xcolor}
\usetikzlibrary{shapes.geometric, arrows.meta, bending}

\definecolor{GreenPlugins}{HTML}{2bb575}
\definecolor{GreyCore}{HTML}{f2f2f2}
\definecolor{GreyData}{HTML}{666666}

\NewDocumentCommand{\inlineCircle}{ O{black} O{} O{6pt} O{12pt}}
{
\kern-2pt
	\resizebox{#3}{!}{
        \begin{tikzpicture}
            \draw[draw=#1, dash pattern=#2, line width=#4](-1,0) circle (1);
        \end{tikzpicture}
	}
\kern-2pt
}

\NewDocumentCommand{\inlineDisk}{ O{white} O{black} O{8pt} O{2pt}}
{
\kern-2pt
	\resizebox{#3}{!}{
        \begin{tikzpicture}
            \filldraw[fill=#1, draw=#2, line width=#4](-1,0) circle (1);
        \end{tikzpicture}
	}
\kern-2pt
}

\newcommand\inlineRect[2][8pt] 
{
\kern-4pt
	\resizebox{#1}{!}{
        \begin{tikzpicture}
            \fill[fill=#2](0.1,0.1) rectangle (0.4,0.4);
        \end{tikzpicture}
	}
\kern-4pt
}

\NewDocumentCommand{\inlineRectangle}{ O{black} O{white} O{8pt} O{2pt}}
{
\kern-4pt
	\resizebox{#3}{!}{
        \begin{tikzpicture}
            \filldraw[draw=#1, fill=#2, line width=#4](0.1,0.1) rectangle (0.4,0.4);
        \end{tikzpicture}
	}
\kern-4pt
}

\NewDocumentCommand{\inlineArrow}{ O{black} }
{
\kern-2pt
    \begin{tikzpicture}
        \draw [draw=#1, arrows = {-Stealth[#1, length=5pt, inset=1pt]}] (0,0) to [bend left] (0.4,0.1);
     \end{tikzpicture}
\kern-2pt
}

\usepackage{xparse}
\usepackage[user,hyperref]{zref}
\makeatletter
\providecommand{\@currentshorttitle}{}
\zref@newprop{shorttitle}{\@currentshorttitle}
\zref@addprop{main}{shorttitle}

\NewDocumentCommand{\labelshort}{om}{%
  \begingroup
  \IfValueT{#1}{%
    \renewcommand{\@currentshorttitle}{#1}%
    \zlabel{#2}%
  }%
  \endgroup
  \label{#2}%
}

\usepackage[resetlabels]{multibib}
\newcites{supplement}{Supplemental references}

\begin{document}

\firstsection{Introduction} \label{sec:intro}

\maketitle

High-dimensional data has become important and ubiquitous in many applications.
Yet, understanding this type of data remains challenging and poses many hurdles ranging from computational efficiency to interpretability.
Combinations of automated analysis and interactive visualizations, visual analytics (VA)~\cite{Cook2005Path, Keim2008VA}, have proven to assist well in gaining insight for high-dimensional data.
A variety of visual encodings and processing algorithms for high-dimensional data exist.
At the same time, specialized application domains require specialized workflows for handling their data and often need to adapt established methods to their use case.
Even though these domains encounter different domain-specific questions, they often deal with similar abstract data set types.
Additionally, abstracting different domain-specific workflows regularly yields similar goals and user tasks~\cite{Brehmer2013TaskTopology, Lam2018BridgingGoalsTasks} which might be tackled with recurring visual encoding components like heatmaps or analytics methods such as dimensionality reduction.
It is time-consuming and wastes development resources to reinvent the wheel by re-implementing, e.g., a linked selection mechanism for multiple coordinated views every time a domain-specific VA solution is needed~\cite{Krueger2019Facetto, Ma2020VaMachineLearning, Sun2020VAPlanning, Pi2021VATraffic, Wu2022VATactics}.
We developed a visual analytics framework, \SoftwareName, as a flexible solution for VA software developers, application designers, and practitioners to implement algorithms and visual encodings, prototype workflow-specific tool sets, and perform their data exploration and analysis respectively.

Existing VA systems for exploring general multivariate data do not meet all of these goals.
Commercial products like Visplore~\cite{Piringer2008Visplore, Piringer2009Visplore} or Spotfire~\cite{Ahlberg1996Spotfire, Ahlberg1994Visual} come with wide feature ranges but are closed-source and not easily extensible.
Older open-source frameworks like \mbox{XmdvTool}~\cite{Ward1994XmdvTool} and GGobi~\cite{Swayne2003GGobi} are mostly limited to visual analysis and lack analytics functions.
ParaView~\cite{Ahrens2005ParaView} and Inviwo~\cite{Jonsson2020Inviwo} are capable of displaying multivariate data as well but focus on field data and the representation of spatial structures.
Business intelligence solutions like Tableau~\cite{Stolte2002Polaris, Tableau} mostly focus on dashboard creation and chart recommendations.
Other fast dashboard prototyping tools, like Keshif~\cite{Yalcin2018Keshif}, provide infrastructure like linked selections of various data visualizations but lack analytics capability.
With \SoftwareName\ we propose a visual analytics framework for general high-dimensional data that is easily extendable and lets both developers and practitioners re-use algorithmic and visualization building blocks for prototyping and reusing visual analytics systems.

Growing data sizes, both in the number of items and dimensions, increasingly complicate interactive analysis.
Progressive visual analytics~\cite{Stolper2014PVA} intends to overcome this issue by continuously providing intermediate results of the current data analysis step.
The ability to control the analysis based on continuous feedback is crucial for progressive VA systems~\cite{Badam2017Steering}. 
In \SoftwareName\ we implement a data-centric and modular framework that facilitates continuous data updates and algorithm steering out of the box.
The \SoftwareName\ core application manages data sets and plugins, which provide both analysis and visualization functionality.
This architecture allows for fast data changes, selection updates, and overall flexible data exploration.
Additionally, since each plugin is agnostic of any other, the system is easy to extend with new data types, visualizations, and analysis algorithms.
\SoftwareName\ is written in C++, using the Qt framework~\cite{Qt} for cross-platform GUI development.
OpenGL is used for high-performance rendering (e.g., our scatterplot plugin) but viewer plugins based on lower threshold JavaScript libraries like D3~\cite{Bostock2011D3} and Vega-Lite~\cite{Satyanarayan2017VegaLite} are also possible. 
\SoftwareName\ is open source and can be found at \href{https://github.com/ManiVaultStudio}{github.com/ManiVaultStudio}. 

\vspace{1.5mm}
To summarize, in this paper we describe 
\begin{itemize}
    \item \SoftwareName, a modular and extensible visual analytics framework designed for high-dimensional data,
    \item several functionality extensions in the form of basic data-, viewer-, and analytics plugins, and
    \item three use cases ranging from plugin development to a practitioner's workflow.
\end{itemize}

\section{Related Work} \label{sec:relatedWork}

Visual analysis of high- and multidimensional data is broadly discussed in literature~\cite{wong1994Years, Fuchs2009VisMulti, Kehrer2013VAsurvey}.
Here, we review the most relevant work on 
Visual Analytics (VA) systems for multidimensional data and
visualization design environments %
with respect to our framework.

\newcommand\tyes{\textbullet}
\newcommand\tno{---}
\newcommand\AccentColor{WhiteSmoke} %

\begin{table*}[ht]
\centering
\begin{threeparttable}\vspace{-3mm}
\caption{\textbf{Comparison with other visual analysis tools} that are most similar to \SoftwareName.\vspace{-2mm}}
\vspace{-3mm}
\label{tab:comparisonTools}
\rowcolors{2}{\AccentColor}{White}  %
\begin{tabular}{lccccccc}
\toprule
 & \SoftwareName & XmdvTool \cite{Ward1994XmdvTool} & GGobi \cite{Swayne2003GGobi} & Visplore~\cite{Visplore} & Tableau \cite{Tableau} & ParaView~\cite{Ahrens2005ParaView} & Inviwo \cite{Jonsson2020Inviwo} \\ 
\cmidrule(r){2-8}
Focus on high-dim. data         & \tyes & \tyes & \tyes  & \tyes  & \tyes  & \tno & \tno  \\ 
Focus on field data     & \tno  & \tno & \tno  & \tno & \tno  & \tyes & \tyes  \\ 
Extensible                      & \tyes & \hphantom{\,$^{\text{a}}$}\tyes\,$^{\text{a}}$ & \tyes & \tno & \tno & \tyes & \tyes  \\ 
Visual Analytics                & \tyes & \tyes & \hphantom{\,$^{\text{b}}$}\tyes\,$^{\text{b}}$ & \tyes & \tyes& \hphantom{\,$^{\text{c}}$}\tyes\,$^{\text{c}}$ & \hphantom{\,$^{\text{d}}$}\tno\,$^{\text{d}}$    \\ 
Progressive Analytics           & \tyes & \tno & \hphantom{\,$^{\text{b}}$}\tyes\,$^{\text{b}}$ & \tyes & \tno & \hphantom{\,$^{\text{d}}$}\tno\,$^{\text{d}}$ & \hphantom{\,$^{\text{d}}$}\tno\,$^{\text{d}}$  \\
VA system authoring             & \tyes & \tno  & \tno  & \tno  & \hphantom{\,$^{\text{d}}$}\tyes\,$^{\text{d}}$  & \hphantom{\,$^{\text{e}}$}\tyes\,$^{\text{e}}$ & \tno \\ 
Active development              & \tyes & \tno  & \tno  & \tyes  & \tyes  & \tyes  & \tyes  \\ 
License                         & LGPL-3 & Public domain & EPL & Commercial & Commercial & BSD-3 & BSD-2 \\ 
\bottomrule
\end{tabular}

\begin{tablenotes}\footnotesize
  \item%
  $^{\text{a}}$~No dynamic extension loading\quad%
  $^{\text{b}}$~When used with its API, e.g., in combination with R\quad%
  $^{\text{c}}$~Via Trame~\cite{trame}\quad%
  $^{\text{d}}$~The systems can be extended with Visual Analytics functionality by plugins or Python integration, but the focus is on interactive field visualization\quad%
  $^{\text{e}}$~Focus on dashboards with pre-populated data
\end{tablenotes}
\vspace{-5mm}
\end{threeparttable}
\end{table*}

\subsection{Visual Analysis and Analytics Systems} \label{sec:relatedWork:VA}
VA systems for the exploration and analysis of high-dimensional data are well established both in academia and industry~\cite{Cui2019OverviewVA, Ghosh2018IndustrySurvey}.
\Cref{tab:comparisonTools} gives an overview comparison between \SoftwareName\ and visual analysis tools that we deem most similar.
Most VA systems employ coordinated multiple views~\cite{Roberts2007StateArtCoordinated} with linked selections for data exploration, and we follow this approach with \SoftwareName\ as well. 
Chen et al.~\cite{Chen2021Composition} discuss common practices and guidelines for the layout of multiple views.

Pioneering visual analysis frameworks for \textit{multidimensional data} include XmdvTool~\cite{Ward1994XmdvTool}, Spotfire~\cite{Ahlberg1996Spotfire}, GGobi~\cite{Swayne2003GGobi} and the InfoVis toolkit~\cite{Fekete2004InfoVisToolkit}.
These frameworks mostly focused on displaying data with a variety of visual idioms and enabled exploration with brushing tools and linked selections. 
XmdvTool was extended with several dimensionality reduction and clustering methods~\cite{Yang2003InteractiveHierarchicalDisplaysXmdvTool, Yang2003VisualHierarchicalDimensionXmdvTool, Cui2006MeasuringDataAbstractionXmdvTool}.
GGobi~\cite{Swayne2003GGobi} integrates with the R language which enables users to apply analysis algorithms via scripting.
Spotfire grew into a commercial, closed-source product with extensive analytics capabilities, while the others are open-source, albeit unmaintained.
All of these tools predate \textit{Progressive} VA and are not optimized for the specific needs of continuous updates and steering of analytics processes.
\SoftwareName\ is designed around the principles of progressive VA from the start using a data-centric architecture.
Data-producing and -transforming plugins can continuously update the data managed by the core, while data consumers get automatically notified about these changes.
Tableau~\cite{Tableau}, building on the Polaris system~\cite{Stolte2002Polaris}, might be the most prominent and representative universal VA system. 
Marketing itself as a business intelligence tool, Tableau focuses on flexible visualization of various data types and more general analytics functions can be added via Python or R scripts.
Similarly, Visplore~\cite{Piringer2008Visplore, Piringer2009Visplore} implements a suit of statistical analysis and visualization methods for tabular data and aims at providing quick visual feedback for visual interactions and data queries.
Its commercial offspring~\cite{Visplore} offers a more direct integration of scripting languages to supplement built-in analysis functions.

The open-source ParaView~\cite{Ahrens2005ParaView}, like many other analysis frameworks for spatial field data, e.g., volume data,~\cite{Bavoil2005VisTrails, 3DSlicer, Childs2012VisIt, Sakamoto2015KVS} is based on the VTK library~\cite{schroeder1998VTK}, and provides a wide range of visualization and analysis functions in an extensible framework.
ParaView follows VTK's visualization pipeline and is designed around the flow of data through various transformations to their final visual presentation.
Similarly, the commercial Amira Software~\cite{Stalling2005Amira, AmiraSoftware} offers a range of analysis functions for multidimensional volumetric data but it is not freely extensible.   
Many visual analysis systems traditionally target either geometric or abstract tabular data. %
However, in recent years, the analysis of spatial and non-spatial data has become increasingly integrated~\cite{Sorger2015Integration}.
With \SoftwareName\ we create a system for general high-dimensional data that can be extended to handle arbitrary spatial or abstract data types.
Our data-centric system design enables flexible exploration workflows instead of having practitioners concerned about data flow through each step of the visualization pipeline.

\subsection{Visualization Design Environments} \label{sec:relatedWork:VDE}
\textit{Visualization design environments} or similarly visualization prototyping systems are tools for creating visualizations that provide a graphical user interface for specifying visual encodings of data and interaction dynamics.
Many such systems exist, and here we provide an overview of the tools most similar to \SoftwareName.

Lyra~\cite{Satyanarayan2014Lyra} offers fine-grained design options for single plots through handles, drop-zones, and other interaction mechanisms for graphical setup of re-usable Vega or Vega-Lite~\cite{Satyanarayan2017VegaLite} specifications.
{Lyra 2}~\cite{Zong2021Lyra2} extends this framework by letting users define interactions like brushing and selection linkage between multiple plots.
iVisDesigner~\cite{Ren2014IVisDesigner} follows similar principles but places emphasis on collections of data visualizations in a dashboard format.
Keshif~\cite{Yalcin2018Keshif} focuses on a novice user audience by automatically aggregating data and selecting visual representations based on pre-defined mappings for various data types.
In contrast to the above design environments for single or multiple visualizations, \SoftwareName\ is a design environment for complete visual analytics systems including automated analysis methods.
While the above systems are focused on abstract data, Inviwo~\cite{Jonsson2020Inviwo} presents a visualization prototyping system for spatial field data.
Its design allows users to specify visualizations on various abstraction levels, from visual (connecting functional boxes) to conventional programming. 
Compared to Inviwo's data-flow model, \SoftwareName\ is data-centric and focused on providing several visualizations and analytics tool building blocks. 
\SoftwareName's core system coordinates views on the data and enables linked selections between views out-of-the-box.

From a plugin-in developer's perspective, \SoftwareName\ resembles the {prefuse}~\cite{Heer2005Prefuse} and ComVis~\cite{Matkovic2008ComVis} toolkits.
They provide development environments and software components for building dynamic visualizations.
Both focus on non-spatial data and target graph and tabular data set types. 
Scripting-based solutions like Dash~\cite{plotlyDash} for creating dashboard applications or Voilà~\cite{kluyver2016jupyter} for converting Jupyter notebooks into standalone web pages provide a GUI front-end to the wide offer of analysis libraries in the Python, R or Julia ecosystems.
\SoftwareName\ is specifically laid out for progressive and high-dimensional data analysis.
Our C++ implementation supports high-performance computations and interactions necessary for visual analytics.

\section{Design Considerations} \label{sec:design}
We designed \SoftwareName\ as a VA framework with multiple user groups in mind. 
While these groups can overlap, their requirements for the effective and convenient use of \SoftwareName\ are varied.

\subsection{General Setting} \label{sec:design:Setting}
High-dimensional data has become ubiquitous in many domains and the analysis of such data plays a pivotal role in acquiring insights into complex systems.
Analytics software in different domains targeted at such data generally utilizes comparable sets of analytical and visual tools, such as dimensionality reduction, clustering algorithms, scatterplots, or parallel coordinates plots. 
These generic tools are then combined with data-, user-, and domain-specific tools and customizations to create a specific application.
The primary motivation for developing \SoftwareName\ is to facilitate rapid construction of visual analytics applications for high-dimensional data without the need to re-implement common functionality. 
Modularity is a key aspect for creating reusable tools, both on a code as well as a user-facing abstraction level.
The second main motivation for \SoftwareName\ is a need for flexible exploratory analysis, but also subsequent sharing of results, as well as the means to recreate the corresponding workflows.
We learned of the target user characteristics and design requirements during multiple collaborations with practitioners in various fields~\cite{Hollt2016Cytosplore, Popa2022Endmember, Thijssen23, Li2023Space} spanning several years.

\subsection{Target Users} \label{sec:design:users}
We identified three target user groups, each with specific requirements:
\begin{enumerate}[label=\textbf{U\arabic*}, wide, labelwidth=0pt, labelindent=0pt]
    \item\label{Usr::Developer} \textbf{Developers} use \SoftwareName\ to implement new ideas and methods. 
    These users, e.g., visualization researchers, interact with the system via code in order to create customized modules. 
    Developers need the framework to provide a stable API that allows for the integration of their methods with little overhead.
    Further, they need existing modules to focus on their specific contribution; e.g., a developer of a dimensionality reduction method might want to visualize results in an existing scatterplot module without having to implement their own.
    \item\label{Usr::Designer} \textbf{Application designers} combine and adapt {existing} modules to create stand-alone applications for specific use-cases.
    Not all options of a view (e.g., the point size in a scatterplot) might be necessary for a specific workflow, and providing all options in the GUI can be distracting.
    In these scenarios, \SoftwareName\ needs to support flexible GUI customization.
    To minimize the burden, the framework should support such customization directly in the GUI without programming.
    \item\label{Usr::Practitioner} \textbf{Practitioners} and domain experts use the software to analyze their high-dimensional data.
    Practitioners need \SoftwareName\ to allow for a flexible data exploration process, to provide responsive user interfaces, and to offer domain-specific visualization and analysis modules.
    Once their analysis is finished, practitioners need the ability to easily share and reproduce the results and their workflow in \SoftwareName. 
    Given a well-defined workflow, they also need easy access to specified presets of visualization and analysis layouts.
\end{enumerate}

The boundaries between these user groups are fluid.
E.g., a skilled practitioner might want to extend a pre-bundled application with a module or develop a module themselves.

\subsection{System Requirements}\label{sec:design:requirements}
Based on the general usage setting and needs of our target users, we define the following high-level requirements for a visual analytics platform such as \SoftwareName.
The framework must be:
\begin{enumerate}[label=\textbf{R\arabic*}, wide, labelwidth=0pt, labelindent=0pt]%
	\item\label{Req::Extensible} \textbf{Extensible}: \SoftwareName\ has to provide an interface for adding new functionalities.
    It must be possible to create modules for new
    \vspace{-1.5mm}
    \begin{enumerate}[label=\textbf{\alph*}, itemsep=0pt, parsep=0pt] 
        \item data types,
        \item visualizations,
        \item analytics methods,
        \item data transformations,
        \item loading/writing data.
    \end{enumerate}
    \vspace{-1.5mm}
    \item\label{Req::Flexible} \textbf{Flexible}: \SoftwareName\ must allow for workflows in multiple domains and specifically enable straightforward workflow adaption during use.
    \item\label{Req::Linkable} \textbf{Linkable}: \SoftwareName\ must provide modules with an API to easily link data selections and synchronize parameters, such that no dependencies between modules are created.
    \item\label{Req::Configurable} \textbf{Configurable}: \SoftwareName\ must provide options for GUI configuration during runtime through the user interface.
    \item\label{Req::Distributable} \textbf{Distributable}: \SoftwareName\ must be able to save its current state, including layout, data sets, and settings and reproduce a saved state.
    \item\label{Req::Performance} \textbf{Performant}: \SoftwareName\ must be performant when handling large data, stay responsive and provide interfaces to interact with processes during calculation to support progressive VA.
\end{enumerate}

\section{ManiVault Architecture} \label{sec:architecture}

\begin{figure*}[t]
    \centering
    \vspace{-1mm}%
    \includegraphics[width=0.95\textwidth]{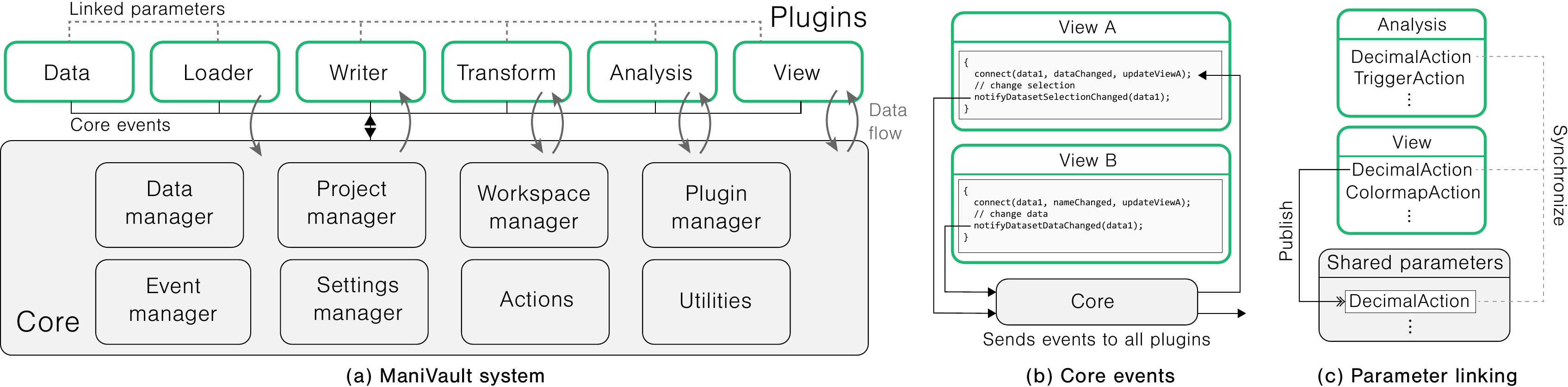}%
    {\phantomsubcaption\ignorespaces\label{fig:overview:system}}%
    {\phantomsubcaption\ignorespaces\label{fig:overview:corevents}}%
    {\phantomsubcaption\ignorespaces\label{fig:overview:paramshare}}%
    \vspace{-1mm}%
    \caption{\textbf{\SoftwareName's system architecture.} The core manages data and events, provides GUI management (actions), etc. Green 
    \inlineDisk[GreenPlugins][GreenPlugins][5pt][1pt]
    borders indicate plugins, a light-grey \inlineDisk[GreyCore][GreyCore][5pt][1pt] 
    background the core. Data flow from the core to data consumer plugins and from data producer plugins to the core is indicated with 
    \inlineArrow[GreyData] 
    arrows. %
    (b) View A listens to \texttt{notifyDatasetDataChanged} emitted by View B. %
    View B does not listen to the \texttt{notifyDatasetSelectionChanged} event triggered by View A, but any plugin could. %
    (c) a view plugin published a DecimalAction, moving the action in a shared parameters space and immediately subscribes to it. 
    Now, an analytics plugin can connect to the shared action, enabling synchronization across plugins. %
    }
    \label{fig:overview}
    \vspace{-5mm}
\end{figure*}

In order to ensure easy extensibility (\ref{Req::Extensible}), \SoftwareName\ is implemented as a modular system, see \autoref{fig:overview:system}.
The core application is a lightweight set of managers and any user-facing functionality is dynamically loaded from self-contained libraries, i.e., plugins, respectively discussed in \Autoref{sec:architecture:core, sec:architecture:plugintypes} (\ref{Req::Performance}).
This compartmentalization into a core and extensions provides easier maintainability, better scalability, and faster development.
Together with a data-centric system structure (\autoref{sec:architecture:data}), this enables flexible workflows (\ref{Req::Flexible}) with various analytics and visualization techniques.
\SoftwareName\ features an intricate notification and parameter sharing system to allow for communicating between plugins, see \autoref{sec:architecture:plugin_communication} (\ref{Req::Linkable}).
GUI management objects, called actions (\autoref{sec:architecture:actions}), implement a part of the communication system and the configuration and serialization system, see \Autoref{sec:architecture:projects, sec:architecture:studiomode} (\ref{Req::Configurable}, \ref{Req::Distributable}).

\subsection{Core Application}\label{sec:architecture:core}
\SoftwareName's core is modularized into a set of managers, actions, and utilities as shown in \autoref{fig:overview:system}.
\SoftwareName\ comprises a data-centric architecture: a data manager stores and administers access to data sets.
All data sets are organized hierarchically, such that derived data sets like clusterings, embeddings, or proper subsets are marked as children of their respective source data.
This enables simple access to properties of the parent data set and propagation of selections from derived to source data sets.
Analysis, transformation, visualization, and loading/writing functionality as well as the definition of data types themselves are separated into plugins.
A plugin manager loads plugins into the core and makes them available to the user.
Each plugin can \textit{consume data}, i.e., process existing data in the core and/or \textit{produce data}, i.e., store a new or alter an existing data set in the core.
While each plugin is self-contained, communication between plugins is made possible using two messaging systems (\autoref{sec:architecture:plugin_communication}).
An event manager in the core administers globally defines notifications while actions are used for run-time configurable notifications (see \Autoref{fig:overview:corevents, fig:overview:paramshare}). 

The general application layout is handled by a workspace manager which takes care of the arrangement of all GUI widgets provided by view plugins.
The core contains two main system view plugins, a data hierarchy, and a data properties viewer. 
The former displays the internal hierarchical data structure, while the latter shows properties of the data (number of data points, dimensions, active selections) and gives users access to the settings of analytics plugins, as discussed in more detail in \autoref{sec:implementation}.
\SoftwareName\ provides a number of actions, GUI management objects, and administers any user-defined linking between them, see \autoref{sec:architecture:actions}.
Further, a project manager is responsible for saving and loading the current state of the application, including loaded data sets, the GUI layout, opened plugins, and linked parameters.
Global settings applicable to, e.g., all plugins or the general application layout are handled by a dedicated settings manager.

Additionally, \SoftwareName's core supplies a set of utilities like dedicated renderers, shaders, color maps, mathematical helper classes, such as vectors and matrices, as well as common algorithms like mean shift clustering.
These tools can be used to create a more coherent visualization and analysis setup across plugins.
E.g., developers can rely on the availability of a standard set of color map types in every view plugin, while maintaining the ability to introduce custom ones.

\subsection{Plugin Types}\label{sec:architecture:plugintypes}
\SoftwareName\ works with six distinct plugin types that bundle various types of functionality.
The system can be easily extended with new functionality by writing a new plugin that will automatically be loaded on start-up (\ref{Req::Extensible}).
In combination with the data-centric core architecture, this enables a user to perform flexible workflow changes (\ref{Req::Flexible}).
\begin{itemize}[wide, labelwidth=0pt, labelindent=0pt]
    \item[\textbf{Data plugins}] enable extending the types of data the system can handle. 
    \SoftwareName\ provides a base data plugin class that developers can extend to define a custom data format. 
    E.g., we provide an image data type that extends our basic point data type with image dimensions and thus a mapping of points to image coordinates. 
    The system can generally be extended with arbitrary data formats.
    \item[\textbf{View plugins}] provide a view on the data and allow interaction, such as selection of data elements. 
    Views can be fully-fledged visualizations or simpler views such as lists.
    View plugins are primarily \textit{data consumers}, i.e., they take a data set as input for visualization, but can also function as \textit{data producers}, e.g., by providing means for annotating data. 
    We provide example plugins with diverse backends, like OpenGL and D3.
    \item[\textbf{Analytics plugins}] allow for the implementation of data analytics modules such as dimensionality reduction. 
    As such, they are primarily \textit{data producers} but also follow the \textit{data consumer} API to receive the input data on which they perform calculations.
    \item[\textbf{Transformation plugins}] resemble analytics plugins in code but are semantically different. 
    They are also primarily \textit{data producers}, but while analytics plugins derive new properties, e.g., an embedding, that can have an arbitrary shape, transformation plugins produce data of the same shape, i.e., with identical items and attributes. 
    An example of such a transformation is a normalization of the original data.
    \item[\textbf{Loader/Writer plugins}] respectively load specific types of data into the system (\textit{data producer}) or write it back to file (\textit{data consumer}).
\end{itemize}

\subsection{Data Handling} \label{sec:architecture:data}
The data handling in \SoftwareName\ follows a model-view pattern.
Internally, the core's data manager keeps a list of raw data models, data set views, and selection views.
A data plugin has to define both a raw data model and data set view \textemdash\ the selection view is simply another instance of the same data set view on the raw data.
The raw data model holds the physical data values of a set and is never exposed directly to non-data plugins.
Therefore, for most intents and purposes, the data set views can be regarded as the actual data sets present in the system.
They define access to the raw data for all non-data plugins by providing, e.g., views on or copies of it.
Each raw data object is associated with exactly one selection object to ensure straightforward selection sharing across all plugins that access a data set.
Selection and set views can be separately requested and adjusted.
This model-view pattern allows for a simple API and to create and use subsets with minimal overhead.

New data sets can be marked as derived from existing ones, e.g., when a new data set is created by an analytics plugin.
The derived data also functions as the user-facing entry point through which the analytics settings can be accessed.
This operation will create new data set and raw data objects but no new selection view.
Instead, selection views are shared between parent and derived data sets.
This simplifies the propagation of selections between views, e.g., a derived embedding shown in a scatterplot and the original data in a parallel coordinates plot.
To enable selection sharing between arbitrary data sets, \SoftwareName\ lets users group data sets in the hierarchy view.
Selections of any data sets within a group and with the same number of data points are then automatically synchronized.

We implemented a set of base data plugins in \SoftwareName{}, including plugins for point data, multichannel images, clusters, color, and text data.
The development of \SoftwareName\ so far primarily targeted the point data type, which can store various high-dimensional integer and floating point formats.
Our image data plugin shows the versatility of \SoftwareName's data handling and the point data type.
When loading an image, two data sets are created: a point data set whose raw data object stores the actual pixel values and a child image data set whose raw data object stores metadata like image size. 
The image data set view provides access to the parent's raw data.
This configuration ensures compatibility with analytics, transformation, and view plugins that expect point data to process multichannel images.

The implemented data handling system is lightweight. Besides the basic \SoftwareName\ core ($<90$ MB), the data manager and hierarchy require $<8$ MB of memory (on Windows). 
Each loaded data set produces less than $1.5$ MB overhead in addition to its binary size, stemming from the plugin instance and core integration.
More details can be found in Supplemental Material S1.

\subsection{Plugin Communication}\label{sec:architecture:plugin_communication}
Coordinated Multiple Views (CMVs)~\cite{Roberts2007StateArtCoordinated} are the basis for virtually any visual analytics application.
While the individual views in a CMV system naturally map to modules in a modular architecture, an essential part of CMV systems is the integration of those views.
This enables techniques like brushing and linking~\cite{Buja1991Linking}, where selections on the data are propagated to all views in the system, or the synchronization of parameters, like the viewport in an Overview+Detail system~\cite{PlaisantOverviewDetail}.
Enabling such linking of views, without breaking the system's modularity (\ref{Req::Linkable}) is no trivial task.
A plugin should be self-contained with respect to its functionality. 
Yet, at the same time, plugins need to be able to communicate, such that they can inform other plugins about data changes and that their parameters can be linked and synchronized throughout the application.

We have designed and implemented two interfaces to solve the issue of inter-plugin communication.
First, an event-based communication API to cover common system-wide types of events related to data set changes (\autoref{sec:architecture:events}) and second a parameter-sharing API (\autoref{sec:architecture:shared_parameters}) as part of our GUI building blocks (\autoref{sec:architecture:actions}).

\subsubsection{Core Events}\label{sec:architecture:events}
The \SoftwareName\ core API provides an event-based system for inter-plugin communication using the \href{https://en.wikipedia.org/wiki/Publish-subscribe_pattern}{publish-subscriber} pattern. 
Plugins send predefined events to the core, which distributes them, and all subscribers (typically plugins) can digest these events as depicted in \autoref{fig:overview:corevents}.
To efficiently support linking and brushing (\ref{Req::Linkable}), we have implemented such events for any changes of data values like addition (\texttt{notifyDatasetAdded}),
updates (\texttt{notifyDatasetDataChanged}), 
removal (\texttt{notifyDatasetRemoved})), 
changes to data selections (\texttt{notifyDatasetSelectionChanged})
and several other data related changes.
A plugin can choose to listen to all events of a certain type or subscribe only to certain events concerning a specific data set.

An example of a linked selection is shown in \autoref{fig:sigmaconnection}. 
The figure shows a screenshot with three views, a scatterplot and a density plot on the left, and the properties of a clustering analysis on the right.
Clicking any cluster in the clusters list (\autoref{fig:sigmaconnection:clusterlist}) will update the selection set attached to the data set and notify the core of these changes with the \texttt{notifyDatasetDataSelectionChanged} event.
The core will then emit the \texttt{dataSelectionChanged} event with the changed data as an argument and subscribed plugins will receive a notification that triggers a refresh of the view with the updated selection (red points in \autoref{fig:sigmaconnection:scatterplot}).

\subsubsection{Shared Parameters}\label{sec:architecture:shared_parameters}
We designed a complementary API to share parameters between modules (\ref{Req::Linkable}) using GUI actions (\autoref{sec:architecture:actions}).
With this system, a plugin parameter is exposed to other plugins by placing it in a public shared parameter pool, i.e., the parameter is \textit{published} (\autoref{fig:overview:paramshare}).
From there, other plugins can \textit{subscribe} to published parameters (provided that the parameter types match).
Any change to a published parameter will be synchronized with all subscribed parameters.
We provide common GUI elements with \SoftwareName, that developers can integrate into their plugins such that the user can publish a parameter or subscribe to any published parameter at run-time through the GUI (\ref{Req::Configurable}).

\begin{figure}[t!]  
    \centering
    \pdftooltip{%
    \includegraphics[width=\linewidth]{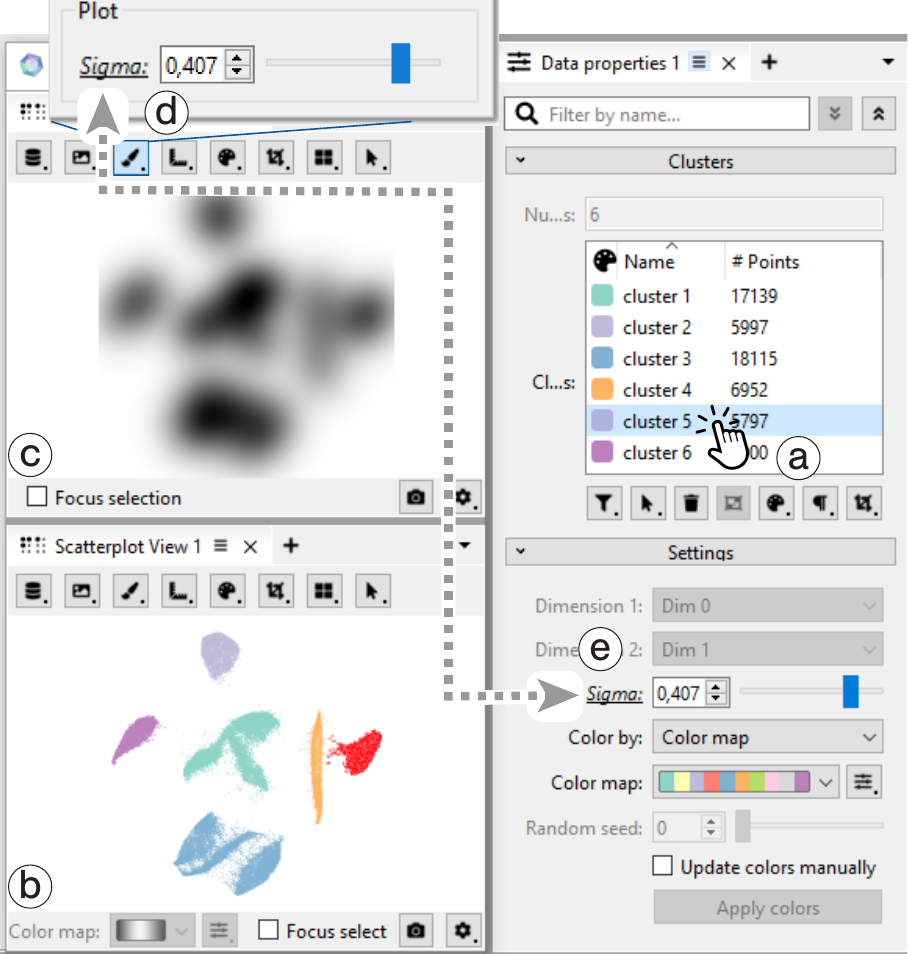}%
    }{%
    Screenshot of a density view of a scatterplot, a scatterplot recolored according to mean-shift clusters and mean-shift analysis properties where a highlighted connection between the sigma parameter of the density visualization and the mean-shift clustering indicates parameter sharing.
    }
    \caption{\textbf{Parameter sharing }by connecting two actions of the same type in the GUI. Both, the Mean-Shift plugin and Scatterplot plugin use a DecimalAction to steer their computation and view respectively.%
    \vspace{-4mm}}    
    \label{fig:sigmaconnection}%
    {\phantomsubcaption\ignorespaces\label{fig:sigmaconnection:clusterlist}}%
    {\phantomsubcaption\ignorespaces\label{fig:sigmaconnection:scatterplot}}%
    {\phantomsubcaption\ignorespaces\label{fig:sigmaconnection:densityplot}}%
    {\phantomsubcaption\ignorespaces\label{fig:sigmaconnection:sigmaplot}}%
    {\phantomsubcaption\ignorespaces\label{fig:sigmaconnection:sigmacluster}}%
\end{figure}

\Cref{fig:sigmaconnection} presents an example in the form of the kernel bandwidth~(sigma) parameter used in kernel density estimation (KDE) employed in density plot visualizations (\autoref{fig:sigmaconnection:densityplot}) but also mean-shift clustering. 
We have implemented plugins for both that allow real-time changes of the sigma parameter, based on Lampe and Hausers real-time KDE~\cite{DaaeLampe2011}.
Linking this parameter between the density plot and the clustering module enables visually finding a suitable density estimation while the clustering is updated on-the-fly.
To link the parameters the user simply clicks on the underlined label in the GUI (\autoref{fig:sigmaconnection:sigmaplot}), e.g., in the density plot view, and chooses "publish".
After defining a suitable name for the parameter, the user can then click on the corresponding label in the settings widget of the mean shift clustering plugin (\autoref{fig:sigmaconnection:sigmacluster}) and click subscribe to be presented with a list of suitable parameters, including the just defined one.
After subscribing, the connection is indicated by the italic font of the \textit{\underline{Sigma}} label.

\subsection{Actions}\label{sec:architecture:actions}
To support sharing of parameters as described above, but also to make it easy to capture the state of a plugin, configure the GUI and unify the look and feel between plugins, we have devised and implemented a number of building blocks we call \emph{actions} on top of the standard Qt GUI widgets.
These include simple actions for \texttt{decimal} and \texttt{integral} values as well as \texttt{strings} but also more complex elements such as \texttt{colors}, \texttt{color maps}, \texttt{file-pickers}, etc..
In addition to those standard GUI elements we implemented a number of custom actions targeting typical VA applications.
These include a general-purpose selection action, that supports different modalities (brushing, rectangle, lasso, etc.) and Boolean combinations (replace, add, remove), and a dimension picker action that provides a consistent way to select one or multiple dimensions of a data set, e.g., to limit the input to a dimensionality reduction plugin.
Although we believe that we provide large coverage of commonly required tasks with the built-in actions, we also provide an API for plugin developers to create custom actions.

By using our actions API, sharing of parameters as described in \autoref{sec:architecture:plugin_communication} is automatically available through the GUI.
In addition, actions can also be attached to data objects, to expose their functionality to other plugins.
A data producer plugin can, e.g., attach an action to trigger a calculation within the plugin.
Other plugins can query these attached actions and provide the corresponding GUI elements within their scope.
We showcase this in our Hierarchical Stochastic Neighbor Embedding (HSNE)~\cite{2016_eurovis_hsne} analytics plugin.
The plugin creates a hierarchical embedding structure that can be refined interactively.
We attach an action for triggering the refinement to the produced embedding data set.
When viewing the embedding in a scatterplot, the scatterplot view plugin exposes the refine action and other attached actions through the context menu.
The user can then trigger the refinement directly from the scatterplot visualization, even though the actual calculation is carried out by the HSNE plugin.

Besides serving as GUI building blocks, we have also implemented support for serialization in the action system.
Each action can be serialized into a \texttt{\href{https://doc.qt.io/qt-6/qvariant.html}{QVariant}} object, including its complete current state, consisting of whether it is active, visible, writable, and the parameter itself.
All actions that belong to a plugin form a hierarchy that can again be serialized into a QVariant object and from there into a JSON object in memory or file on disk.
As such, a plugin that has consistently been implemented with the actions API supports saving and loading of the state out-of-the-box.
Currently, we use this to create presets of a plugin's configuration and to save the complete state of the application to a project file.
In the future, we intend to extend this to a complete provenance mechanism.

An example of a simple decimal action is the implementation of the Sigma parameter discussed above and shown in \autoref{fig:sigmaconnection:sigmaplot}.
The GUI for this parameter consists of the label, a spinbox, and a slider.
Rather than manually creating the GUI elements, the desired elements can be specified when creating the action.
An example of a customization that we integrated in the decimal action is to show a spinbox or slider individually or both, as in this example.
The action then creates the GUI elements on-the-fly and also makes sure they are synchronized by creating them as linked views on the parameter itself.
The underlined label indicates that the parameter is publishable and/or ready to subscribe, while the italics font indicates that it is already linked.
Clicking the label opens a GUI interface for setting up parameter linking.

\subsection{Projects and Workspaces}\label{sec:architecture:projects}
To save the entire state of the application and fully restore it at a later point in time \SoftwareName\ uses projects (\ref{Req::Distributable}).
Projects extend the serialization of actions, described in \autoref{sec:architecture:actions}, to the core framework, capturing settings and the layout of the CMV system.
In addition, a project contains a complete snapshot of the data hierarchy.
We implemented projects as self-contained, compressed archives that are a combination of human-readable JSON files and binary files. 
Two JSON files are used to save the entire state of the application.
A \texttt{workspace.json} contains the CMV layout and actions state and a \texttt{project.json} saves the data hierarchy and additional project metadata. 
The actual data sets are saved as raw binary blobs, with unique identifiers referenced in \texttt{project.json}, to minimize load and save times.
As such, a project is completely self-contained and can be easily distributed to share findings or simply used to come back to an analysis at a later point in time.

We split the description of the project into \texttt{project.json} and \texttt{workspace.json} to add an additional feature, i.e., the definition of user-defined workspaces.
As described above, the workspace contains the complete spatial arrangement of views (layout configuration) and their complete state. 
A workspace is used to set up a complete tailor-made CMV VA application, including customized GUI elements, but without preset data, as a project would.
To enable easy tailoring of layouts and cross-plugin connections directly in the application, even without programming, we designed the \textit{Studio Mode} for \SoftwareName{}.

\subsection{Studio Mode}\label{sec:architecture:studiomode}
\begin{figure}[t!] 
    \centering
    \pdftooltip{%
    \includegraphics[width=\linewidth]{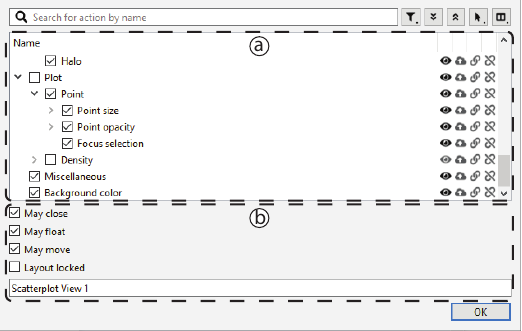}%
    }{%
    Screenshot of a plugin's GUI configuration editor where a designer can toggle action properties on and off (e.g. to disable an action, hide it, or toggle publishing, connecting, and disconnecting on and off).
    }
    \caption{Example of the plugin \textbf{GUI configuration editor} which allows application designers to edit the properties of the plugin actions hierarchy from within the application.
    }    
    \label{fig:EditPluginActions}%
    {\phantomsubcaption\ignorespaces\label{fig:EditPluginActions:hierarchy}}%
    {\phantomsubcaption\ignorespaces\label{fig:EditPluginActions:general}}%
    \vspace{-15pt}
\end{figure}

For the configuration of actions, workspaces, and complete projects, \SoftwareName\ can be put into \emph{Studio Mode}.
This mode of operation allows application designers to create complete tailor-made applications and data viewers from within the GUI of \SoftwareName\ itself.

A plugin editor, shown in \autoref{fig:EditPluginActions}, enables fine-grained control over the user interface.
It lists an overview of all actions that are currently available for opened plugins (\autoref{fig:EditPluginActions:hierarchy}).
Therein each action can be enabled or disabled as a whole \faIcon[regular]{check-square}, but also customized with respect to its visibility \faEye\ or whether it can be published \faCloudUpload*, connected \faLink, or disconnected \faUnlink.
Additionally, the editor lets a user configure general options like the name of a plugin instance, shown in its title bar, or whether the GUI of the plugin may be moved or closed (\autoref{fig:EditPluginActions:general}).

The plugin editor is an essential tool for application designers, to create a completely customized user experience for a specific application.
At the same time, it provides the possibility for advanced users of the system to create presets of views.
Besides saving a complete project, users can adjust the interface of an individual plugin to their needs and save the resulting configuration as a template for future instances of that plugin.
Using the serialization described above, these templates can be saved to disk, providing persistent access across sessions.

For a user-definable flexible layout of the application, we incorporate the Qt advanced docking system~\cite{AdvancedDockingQt} into \SoftwareName{}.
The system allows users and application designers to re-arrange the entire layout according to their needs and preferences.

\begin{figure}[b]
    \centering
    \vspace{-3mm}
    \pdftooltip{%
    \includegraphics[width=\linewidth]{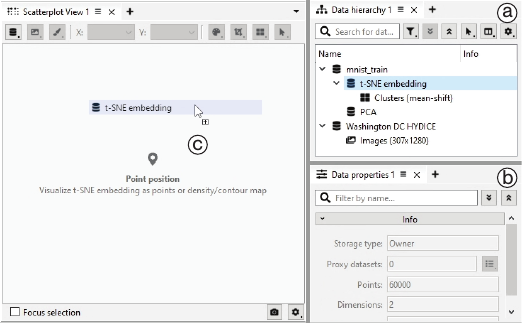}%
    }{%
    A Screenshot of the ManiVault core application, the data hierarchy and properties plugins where a t-SNE embedding data set is dragged from the data hierarchy onto a scatterplot view.
    }
    \caption{\textbf{Data hierarchy} (a) and \textbf{data properties} view (b) in \SoftwareName. Data sets can easily be shown in views via drag and drop (c).
    }    
    \label{fig:core}%
    {\phantomsubcaption\ignorespaces\label{fig:core:hierarchy}}%
    {\phantomsubcaption\ignorespaces\label{fig:core:properties}}%
    {\phantomsubcaption\ignorespaces\label{fig:core:hierarchydrag}}%
\end{figure}

\section{\SoftwareName\ Implementation} \label{sec:implementation}

The \SoftwareName\ core is implemented in C++ and the Qt~\cite{Qt} cross-platform application development framework.
\SoftwareName\ provides a plugin API for data types, view, analytics, transformation, and writer/loader modules.
For each of these types we provide template implementations to lower the entry barrier for developers.
In addition, we have already implemented a number of plugins for various use cases, including some of the core functionality of \SoftwareName\ such as the basic data types, and the data hierarchy and data properties view plugins.

\begin{figure*}[t!]
\centering
\begin{subfigure}{0.195\textwidth}
    \pdftooltip{%
    \includegraphics[width=\textwidth]{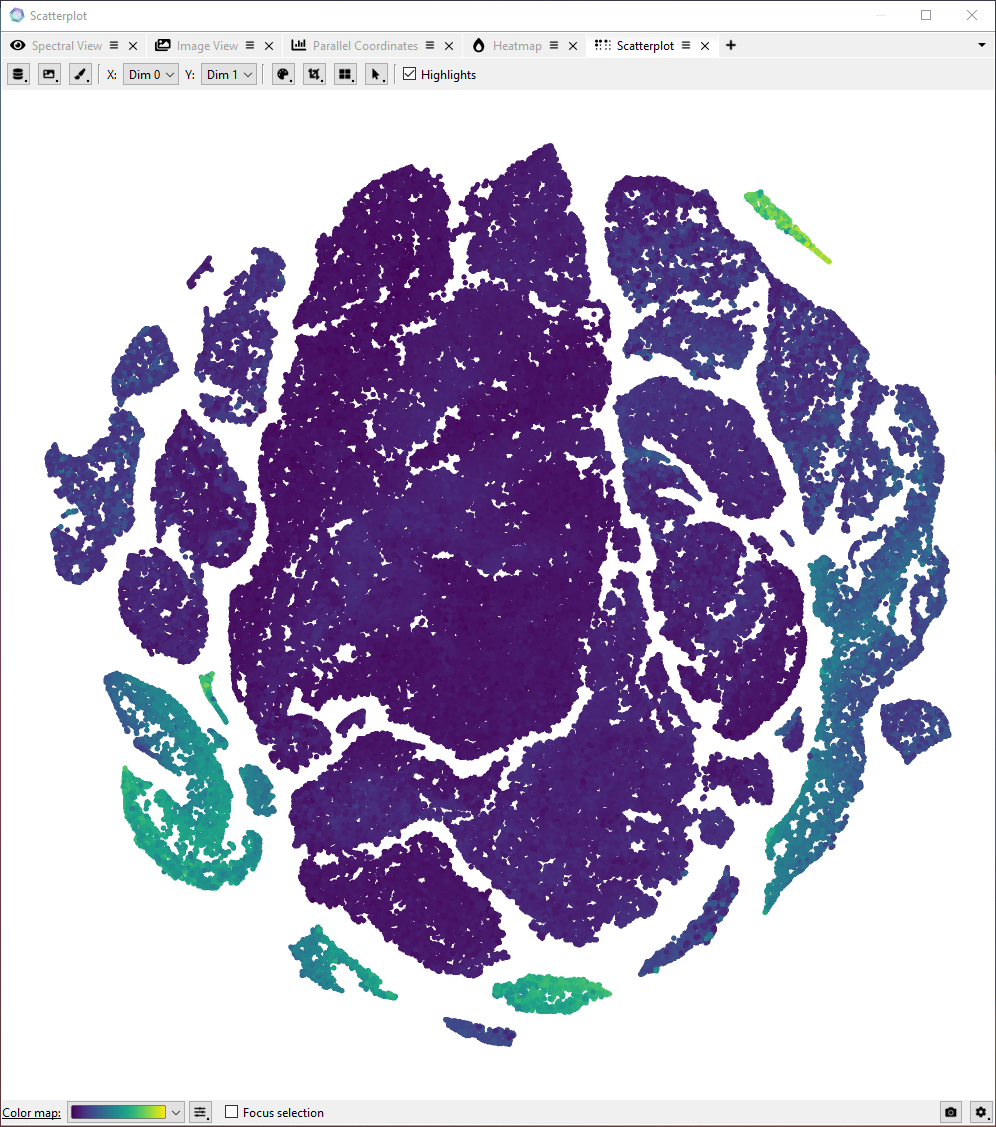}
    }{%
     A Screenshot of an instance of the scatterplot plugin showing a t-SNE embedding of a large hyperspectral imaging data set where points are colored according to one of the spectral bands (dimensions) using the viridis colormap.
    }%
    \caption{Scatterplot}
    \label{fig:viewplugins:scatterplot}
\end{subfigure}
\hfill
\begin{subfigure}{0.195\textwidth}
    \pdftooltip{%
    \includegraphics[width=\textwidth]{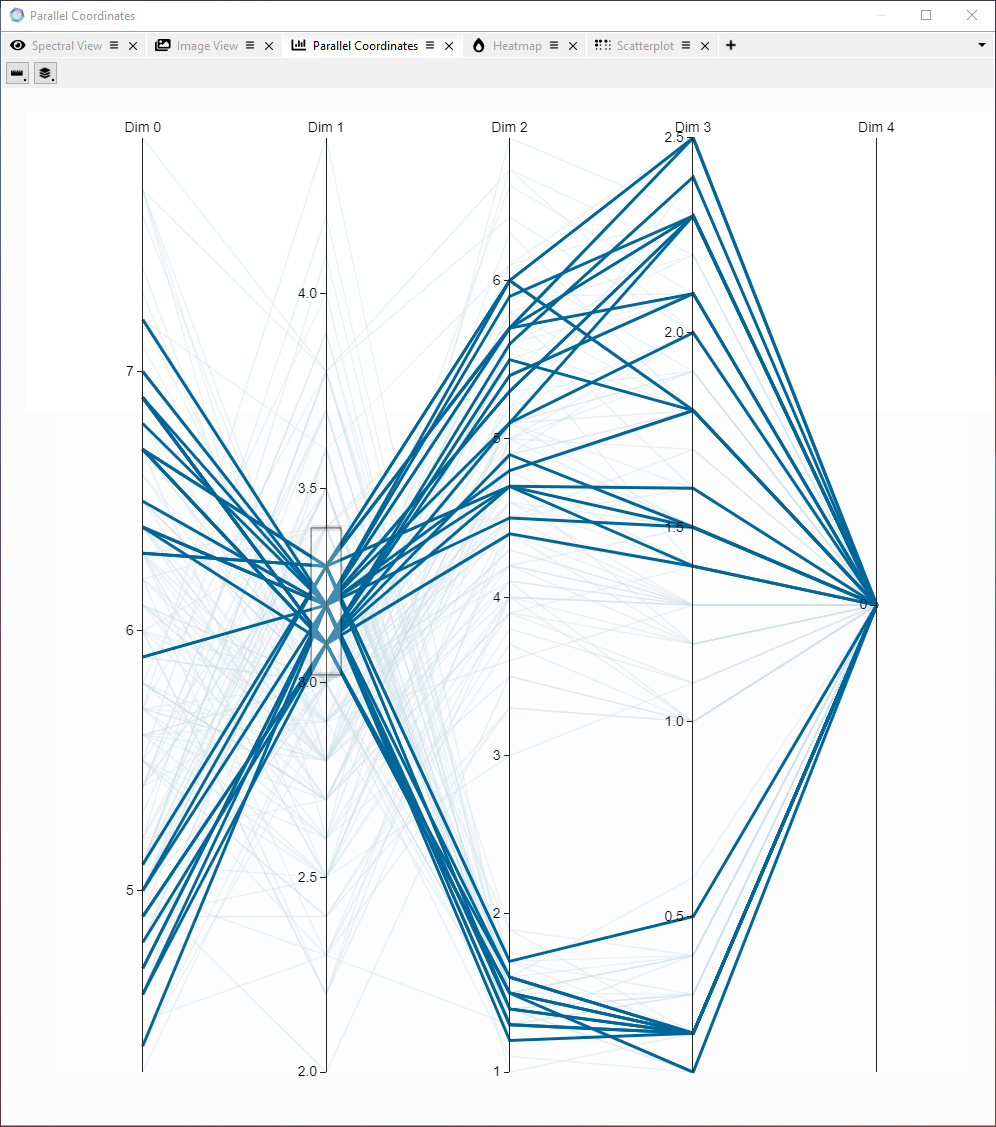}
    }{%
     A Screenshot of an instance of the parallel coordinates plot showing a data set with five dimensions mapping where some of the lines are selected and thus shown in dark blue, while the rest of the lines are shown with high transparency and thus fade into the background.
    }%
    \caption{Parallel Coordinates}
    \label{fig:viewplugins:pcp}
\end{subfigure}
\hfill
\begin{subfigure}{0.195\textwidth}
    \pdftooltip{%
    \includegraphics[width=\textwidth]{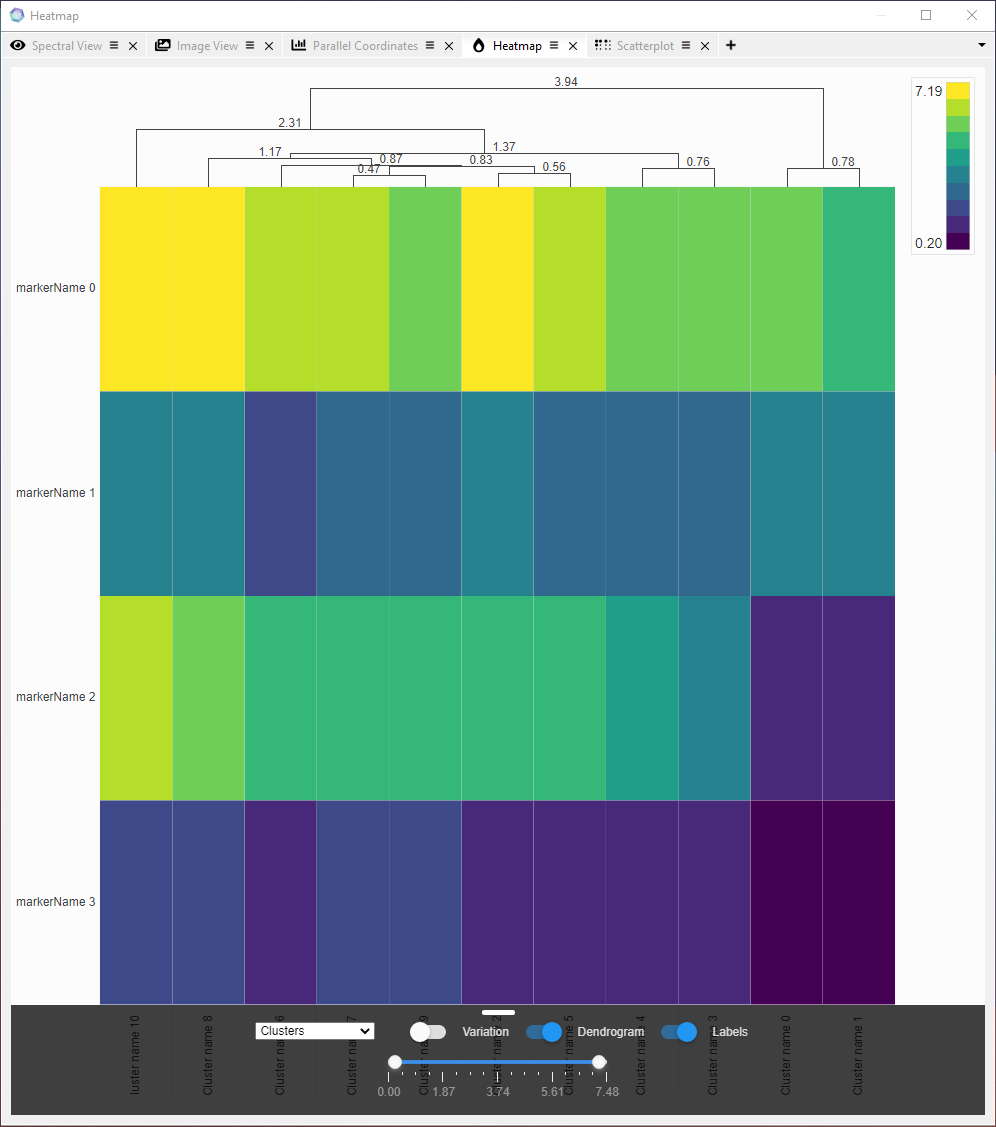}
    }{%
     A Screenshot of an instance of the heatmap plugin showing 11 clusters (mapped to columns) of a data set with four dimensions (mapped to rows) where the median value for each dimension and within one of the clusters is mapped to color using the viridis colormap.
    }%
    \caption{Cluster Heatmap}
    \label{fig:viewplugins:heatmap}
\end{subfigure}
\hfill
\begin{subfigure}{0.195\textwidth}
    \pdftooltip{%
    \includegraphics[width=\textwidth]{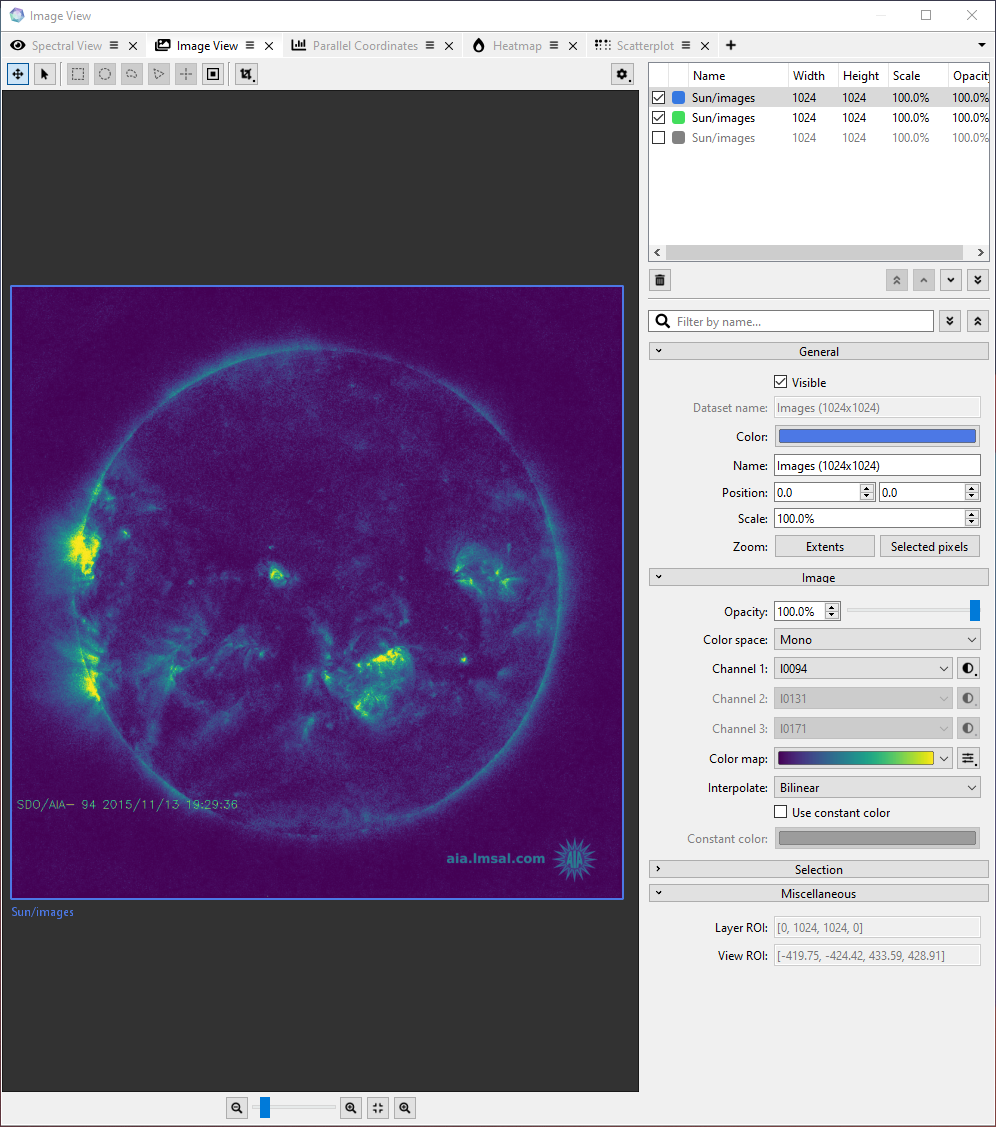}
    }{%
     A Screenshot of an instance of the image viewer plugin showing a hyperspectral image of the sun where pixels are color with respect to one of the reflectance bands using the viridis colormap.
    }%
    \caption{Image View}
    \label{fig:viewplugins:image}
\end{subfigure}
\hfill
\begin{subfigure}{0.195\textwidth}
    \pdftooltip{%
    \includegraphics[width=\textwidth]{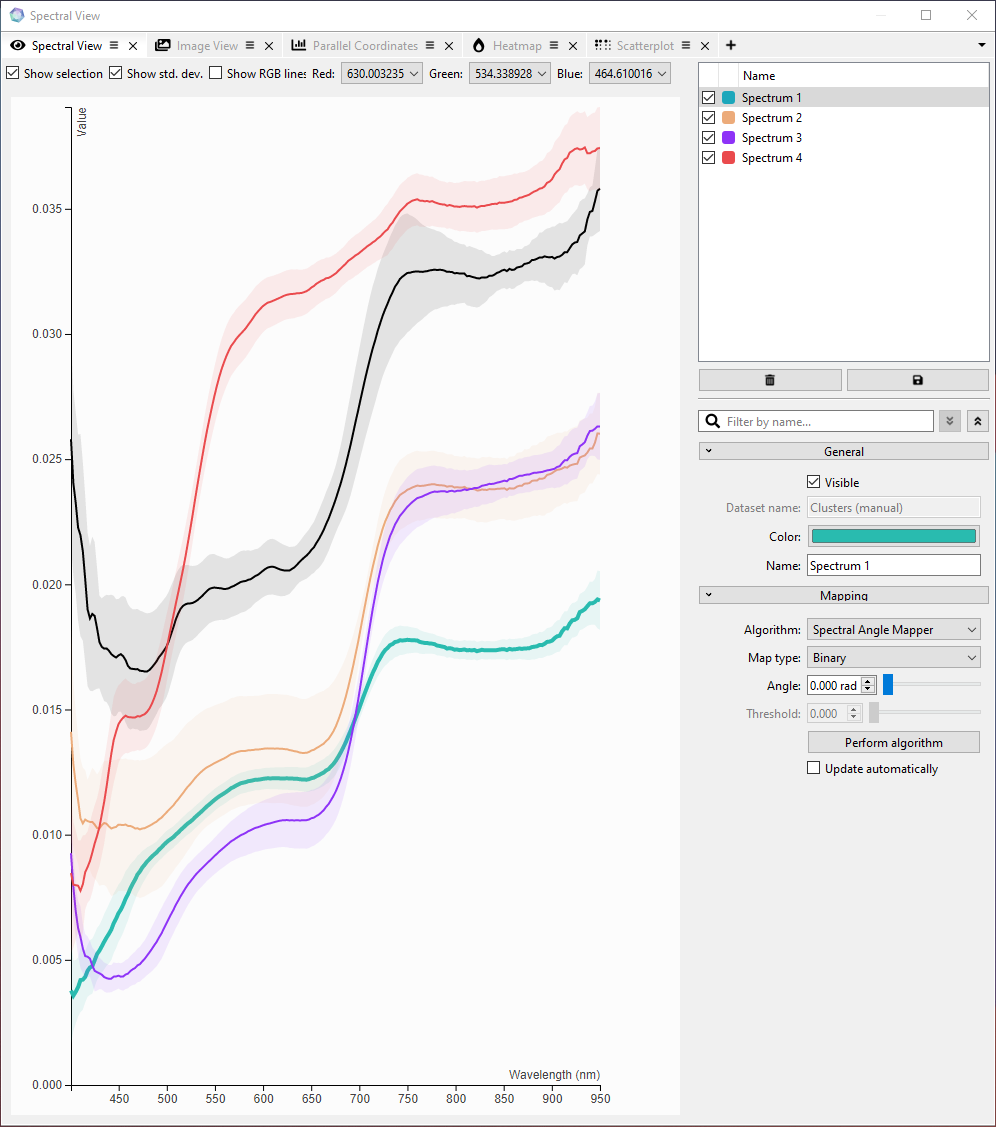}
    }{%
     A Screenshot of an instance of the spectral viewer plugin showing the mean spectra for four clusters as colored lines while a selection (made in another plugin) is shown in black where the standard deviation for all lines is mapped onto a band encompassing the lines.
    }%
    \caption{Spectral View}
    \label{fig:viewplugins:spectral}
\end{subfigure}
\vspace{-3mm}
\caption{A selection of \textbf{viewer plugins} in \SoftwareName.\label{fig:viewplugins}}
\vspace{-5mm}
\end{figure*}

The \textbf{data hierarchy view} (\autoref{fig:core:hierarchy}) functions as the central access point to any data loaded or created in \SoftwareName. 
It displays the data hierarchy in a searchable tree widget where derived data, such as a clustering, are added as children to the original data.
A data set can be loaded into a viewer plugin by simply dragging it from the hierarchy onto the view (\autoref{fig:core:hierarchydrag}).
Alternatively, the user can also interact with each data set through a context menu providing access to all compatible data consumer plugins.
For a fast setup of plugins that expect more than a single input, users can select multiple data sets in the hierarchy and open them through the same menu.
The info panel shows additional information like an analytics progress bar, status messages from plugins or data group affiliation.
If a data set is associated with an analytics plugin, selecting the hierarchy entry will open the analytics settings in the properties view.

The \textbf{data properties view} (\autoref{fig:core:properties}) provides information for a data set selected in the data hierarchy.
For a loaded data set this can be additional metadata created by the loader, e.g., the extents of an image data set.
More importantly, the data properties view also functions as the user interface for analytics and transformation plugins.
These plugins are instantiated through the context menu of a data set, which then functions as their input; their output data sets are then created as children of the input.
Selecting an output data set provides access to the parameters of the analytics or transformation plugin.
\autoref{fig:core:properties} shows the data properties view of an embedding data set, created with our t-SNE plugin.
From here, the user can at any time interact with the t-SNE algorithm, e.g., to pause the calculation, change parameters or compute more iterations.

The data hierarchy and data properties views are integral parts of the system.
More specific functionality is implemented in a number of further plugins.
Dimensionality reduction, integral to high-dimensional data analysis, is provided by 
Principal Component Analysis (\textbf{PCA}), 
t-distributed Stochastic Neighbor Embedding (\textbf{t-SNE})~\cite{vdmaaten2008tsne}, and
Hierarchical Stochastic Neighbor Embedding (\textbf{HSNE})~\cite{2016_eurovis_hsne} plugins.
The t-SNE and HSNE plugins wrap the high-performance HDI library~\cite{pezzotihdi} and as such scale to millions of data points using its GPU-based implementations~\cite{Pezzotti2019Gpgpu}.
For clustering, we provide an interactive \textbf{mean-shift clustering} plugin, based on real-time kernel density estimation~\cite{DaaeLampe2011}.

For visualization, we provide a number of plugins for common plots, including a \textbf{scatterplot} (\autoref{fig:viewplugins:scatterplot}), \textbf{parallel coordinates plot} (\autoref{fig:viewplugins:pcp}), and \textbf{cluster heatmap} (\autoref{fig:viewplugins:heatmap}).
If performance is not a major concern, developers can use web views in combination with \href{https://doc.qt.io/qt-6/qtwebchannel-javascript.html}{Qt's webchannel API} for communicating between the C++ back-end and web-technology-based front-end.
This allows for easily integrating the vast amount of available visualizations in languages like D3~\cite{Bostock2011D3} and Vega-lite~\cite{Satyanarayan2017VegaLite}.
Our heatmap and parallel coordinates plot are based on this technology.
While the webchannel introduces some overhead, such plugins are generally limited by the performance of the JavaScript rendering libraries.
If the scalability of a visualization is of high priority, developers can implement custom high-performance views, e.g., using OpenGL.
We have done so with our scatterplot and image view (\autoref{sec:hdimaging}) plugins.
The scatterplot enables visualization and interaction with millions of points in real-time.
In the default point rendering mode, the different visual channels (point size, color, opacity, etc.) are fully configurable either using fixed values or based on any fitting data available.
Additionally, we implemented a density representation, to provide more visual scalability.

Finally, for data loading and writing, we currently provide support for basic formats in the form of a comma-separated value (\textbf{CSV}) loader/writer and a \textbf{binary} loader/writer.

\subsection{High-Dimensional Imaging}\label{sec:hdimaging}
 
Besides traditional abstract high-dimensional data analytics, we target a number of applications related to high-dimensional imaging (e.g., the workflow presented in \autoref{sec:examples:practitioner}).
As such, we developed a number of plugins targeting such image data.

Central to these efforts is the \textbf{image data type} plugin.
The image data type extends the point data type by the extent of the image.
Consequently, the image data type is compatible with all data consumer plugins that take point data as input; e.g., this allows to calculate a t-SNE using the pixels of a high-dimensional image as input.

We implemented a sophisticated \textbf{image view} plugin (\autoref{fig:viewplugins:image}).
Inspired by widely used image editors, we opted for a layer-based approach.
Users can simply drag multiple data sets into the view, where they are added as layers.
From here, users can define the transparency, as well as the position of each layer, e.g., to stack multiple properties of a single data set as semi-transparent layers or arrange complementing data sets next to each other.
These interactions are possible through standard navigation tools for zooming and panning, while selection is implemented using the action described in \autoref{sec:architecture:actions}.
The actual visualization of the image is fully configurable:
One or two attributes can be displayed by using 1D and 2D color mapping, and three attributes by directly mapping them to the three channels of \emph{RGB}, \emph{HSL}, or \emph{CIELAB} color spaces.

Next to the image viewer, we also provide a \textbf{spectral view} plugin (\autoref{fig:viewplugins:spectral}), specifically for hyperspectral images.
The viewer is based on a simple D3 line plot and shows spectra of individual pixels or, in the case of groups (e.g., selections or clusters), a mean spectrum and a variation as a band around it.

To load image data into \SoftwareName, we currently provide two options.
The first one is a versatile general \textbf{image loader} plugin.
Hyperspectral image data is commonly available as a stack of grayscale images, where each image represents a specific wavelength, also interpreted as a dimension of a high-dimensional space.
Our image loader detects such stack in a folder containing common image formats (including .png, .jpg, .tiff), and also allows direct loading of other common image formats (grayscale, RGB, ARGB). 
Dimensions can be interactively included or excluded from the data set in the loading menu.
We also support re-sampling of the data before loading and the creation of image pyramids to
enable analysis at varying levels of detail, depending on the features of interest or time available for the analysis.
Specific to hyperspectral images, we also provide an \textbf{ENVI loader} plugin compatible with L3Harris' geospatial analysis software ENVI~\cite{ENVI}.

\section{Application Examples} \label{sec:examples}

\SoftwareName\ has already been used for several projects across four universities and several partners.
Popa et al.~\cite{Popa2022Endmember}  and Li et al.~\cite{Li2023Space} describe the design of complete VA systems for analysis of cultural heritage and biological data, respectively. Vieth et al.~\cite{Vieth2022Spidr} and Thijssen et al.~\cite{Thijssen23} developed VA approaches for dimensionality reduction and explaining projections as \SoftwareName\ plugins.
Here, we walk through exemplary usage scenarios for our framework from the perspective of our three target user groups (\autoref{sec:design:users}): software developers (\autoref{sec:examples:developer}), practitioners (\autoref{sec:examples:practitioner}) and application designers (\autoref{sec:examples:applicationdesign}).

\subsection{Writing \SoftwareName\ Plugins -- Developer Perspective}  \label{sec:examples:developer}
\SoftwareName\ provides developers of VA modules with a comprehensive API for data set access, the event notification system, and the other core managers (\autoref{sec:architecture:core}).
Extending the functionality of \SoftwareName\ through new plugins thus comes with minimal overhead.
Example code for each plugin type is available at \href{https://www.github.com/ManiVaultStudio/ExamplePlugins}{github.com/ManiVaultStudio/ExamplePlugins}.

\begin{figure}    
    \centering
    \pdftooltip{%
    \includegraphics[width=\linewidth]{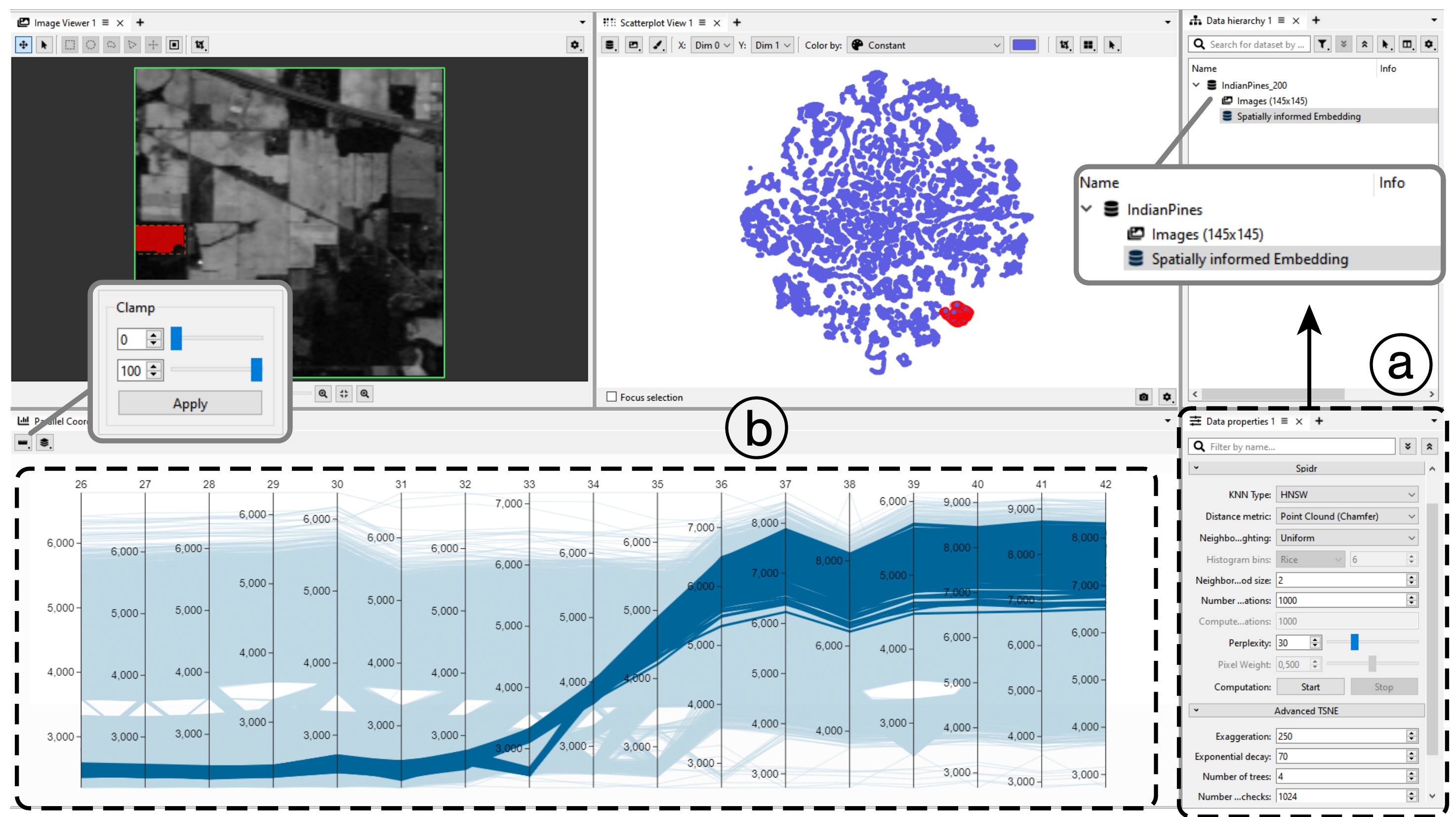}%
    }{%
    Screenshot of ManiVault with three viewer plugins, one analysis, and the data hierarchy view showcasing the interplay of the two plugins implemented in section 6.1 with already available plugins.
    }%
    {\phantomsubcaption\ignorespaces\label{fig:example:SpidrLib:a}}%
    {\phantomsubcaption\ignorespaces\label{fig:example:SpidrLib:b}}%
    \caption{The \textbf{Spidr analysis and parallel coordinates plot} as implemented with the plugin setups from \Autoref{fig:example:pseudoAnalysis, fig:example:pseudoViewer}.
    }    
    \label{fig:example:SpidrLib}%
    \vspace{-15pt}
\end{figure}

\begin{figure}[b]
\begin{lstlisting}[style=CppStyle]
(*@\color{codepurple}void \color{black}AnalyticsPlugin::init() \{ @*)
  (*@\color{codegreen}// 1.\;Derive output from input data set @*)
  (*@\color{black}setOutputDataset(\_core->createDerivedDataset(\color{codered}"outData"\color{black})); @*)
  (*@\color{codegreen}// 2.\;Add settings actions to output data set @*)
  (*@\color{black}outDataset->addAction(\_settings->getSettings()); @*)
  (*@\color{codegreen}// 3.\;Connect GUI interactions (e.g.\:button press) @*)
  (*@\color{codegreen}// \;\;\;\;\;and library callbacks (e.g.\:progress or finish) @*)
  (*@\color{black}connect(\_settings->getStart(), press, \color{codepurple}this\color{black}, runTask); @*)
  (*@\color{black}connect(\_lib, finishedTask, \color{codepurple}this\color{black}, updateCore); @*)
(*@\color{black}\} @*)
\end{lstlisting}
\vspace{-10pt}
\caption{Bare bone \textbf{analytics plugin setup} for wrapping a C++ library. Notifying of output data change (\texttt{step 4}) can be called progressively during the calculation of or on finishing a task.}
\label{fig:example:pseudoAnalysis}
\end{figure}

Here, we present two examples of the necessary steps for creating basic plugins (\ref{Req::Extensible}). 
First, we create an  analytics plugin based on the high-performance t-SNE library HDI~\cite{pezzotihdi}.
In addition, we discuss the implementation of a parallel coordinates plot (PCP) plugin using an existing D3 implementation.
Together with the existing image viewer and scatterplot, these plugins combine into a complete GUI-based application shown in \cref{fig:example:SpidrLib} that is usable by domain expert users without programming knowledge.

To implement the analytics plugin, we follow the steps laid out in \autoref{fig:example:pseudoAnalysis}.
In \texttt{step 1}, we create the output data set by deriving a new data set from the input data, for which the plugin is opened in \SoftwareName.
In this case, we will create a two-dimensional t-SNE embedding containing x- and y-coordinates for all of the points in the input data set.
As such, the output data set will be a points data set that has the same number of points and two dimensions.
Next, we add a settings action to the created data set and define GUI elements using \SoftwareName's action system.
The actions are added to the output data and listed in the data properties view as shown in \autoref{fig:example:SpidrLib:a} (\texttt{step~2}).
We create \texttt{TriggerAction}s which add pushbuttons to the GUI, to start, pause, and resume the calculations and a number of categorical \texttt{OptionAction}s and numerical \texttt{DecimalAction}s, e.g., to expose t-SNE parameters like the distance metric (\texttt{OptionAction}) or perplexity (\texttt{DecimalAction}) (\ref{Req::Configurable}).
Finally, in \texttt{step 3}, calls and reactions to library functions need to be defined.
Here, we notify the core and thereby other plugins about updated output data, in particular, as the t-SNE optimization iteratively progresses, we notify the core after every iteration, such that the viewer plugins can show the progress live.
The result is a lightweight wrapper with no notable performance overhead.
Comparing the performance to running the HDI library using its own Python wrapper showed no performance regression (Supplemental Material S1), even when including progressive updates in \SoftwareName.

\begin{figure}[t]
\begin{lstlisting}[style=CppStyle]
(*@\color{codegray}[ViewWidget.cpp] @*)
(*@\color{black}ViewWidget::ViewWidget() :\!\!\! WebWidget() \{ @*)
  (*@\color{codegreen}// 1.\;Init resources and communication bridge @*)
  (*@\color{codepurple}Q\_INIT\_RESOURCE\color{black}(pcp); @*)
  (*@\color{black}init(\_comObj); @*)
(*@\color{black}\} @*)
(*@\vspace{-6pt}@*)
(*@\color{codegray}[ViewPlugin.cpp] @*)
(*@\color{codepurple}void \color{black}ViewPlugin::init() \{ @*)
  (*@\color{codegreen}// 2.\;Init web widget (set HTML contents) @*)
  (*@\color{black}viewWidget->setPage(\color{codered}":res/pcp.html"\color{black}, \color{codered}"qrc:/res/"\color{black}); @*)
  (*@\color{black}layout->addWidget(viewWidget); @*)
(*@\color{black}\} @*)
(*@\vspace{-6pt}@*)
(*@\color{codegray}[CommunicationObject.h] @*)
(*@\color{codepurple}class \color{black}ComObj :\!\!\! \color{codepurple}public \color{black}WebCommunicationObject \{ @*)
  (*@\color{codegreen}// 3.\;Init signals for communication from cpp to js @*)
  signals:
    (*@\color{codepurple}void \color{black}setData(QVariantList\& data); @*)
  (*@\color{codegreen}// 5.\;Init slots for communication from js to cpp @*)
  public slots:
    (*@\color{codepurple}void \color{black}updateSelection(QVariantList\& selection); @*)
(*@\color{black}\} @*)
(*@\vspace{-6pt}@*)
(*@\color{codegray}[qwebchannel.tools.js] @*)
(*@\color{codegreen}// 4.\;Register signals sent by the view widget@*)
(*@\color{black}bridge.setData.connect(function()\{initPlot(arguments[0])\}) @*)
\end{lstlisting}
\vspace{-10pt}
\caption{Bare bone \textbf{viewer plugin setup} for wrapping a JavaScript library. %
Some boilerplate code is left out for brevity; complete implementation is available alongside other example plugins online.%
\vspace{-5mm}}
\label{fig:example:pseudoViewer}
\end{figure}

To implement the PCP viewer plugin, we need to set up a view widget that shows the PCP chart in addition to settings, like with the analytics plugin. 
Here, the settings are displayed in the same windows as the view widget (\autoref{fig:example:SpidrLib:b}).
Since we build a JavaScript-driven plot, we derive this widget from \texttt{\SoftwareName::WebWidget} and introduce all HTML and JavaScript resources that are used for the PCP through a Qt resource file, \texttt{pcp.qrc} (\texttt{step 1}, \cref{fig:example:pseudoViewer}).
\texttt{Step 2} is to simply set the existing \texttt{pcp.html} file in the existing \texttt{viewWidget}. 
All JavaScript resources are automatically included through the HTML file.
At this point, the viewer is only able to show the content of the provided HTML page.
To establish interactions to and from the C++ side, we set up a \texttt{\SoftwareName::WebCommunicationObject}, which uses a \href{https://doc.qt.io/qt-6/qwebchannel.html}{QWebChannel}.
Within this communication object, we define signals and slots for communication.
E.g., the \texttt{setData} signal (\texttt{step 3}) is used to send the data, provided as a \texttt{QVariantList} object, to a receiver on the JavaScript side.
This receiver, i.e., the \texttt{initPlot} function  is connected in \texttt{step 4} to receive the signal.
Vice versa, slots defined in the communication object can be called directly in JavaScript code, e.g., here we define an \texttt{updateSelection} slot, that can be called from the JavaScript side with a list of selected items.
The plugin then handles any related computations in the corresponding C++ function.

\subsection{Data Exploration -- Practitioner Perspective}  \label{sec:examples:practitioner}
Practitioners in various disciplines work with high-dimensional data sets. 
Here, we consider the exemplary case of exploring remote sensing data using \SoftwareName.
Similar to other application areas, visual exploration of geospatial data is considered important but challenging~\cite{Gahegan2009ExplorationGeo}. 
While specific considerations and final insights will differ from domain to domain, we can follow the task abstraction by Lam et al.~\cite{Lam2018BridgingGoalsTasks} to create a partial workflow that will be representative of many fields (\ref{Req::Flexible}).

\begin{figure*}    
    \centering
    \pdftooltip{%
    \includegraphics[width=\linewidth]{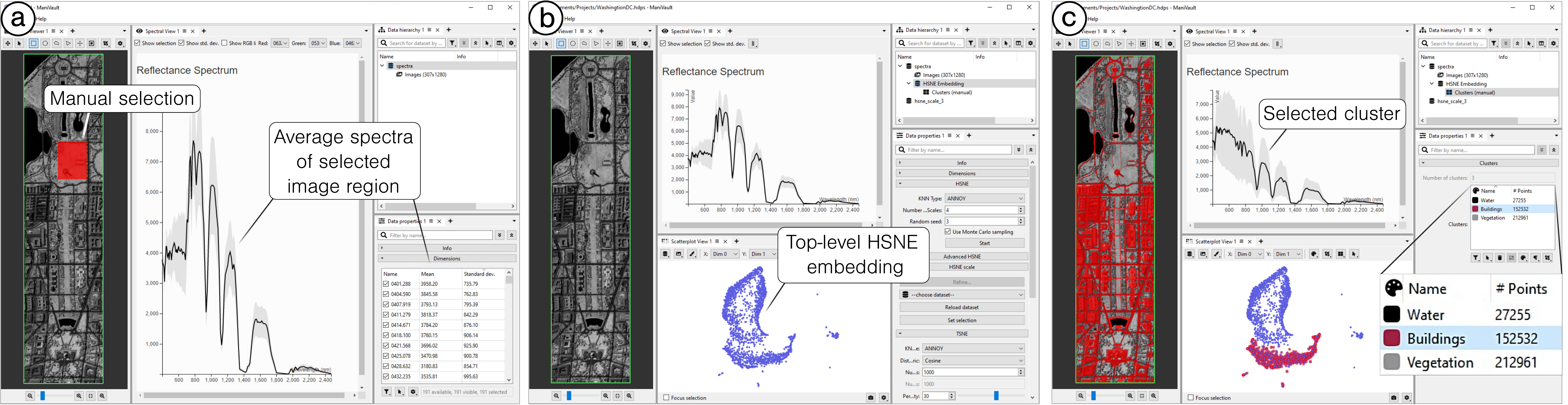}%
    }{%
    Several screenshots of ManiVault showing a typical workflow progression where each screenshot shows the next steps in a data exploration session.
    }%
    {\phantomsubcaption\ignorespaces\label{fig:example:WashingtonWorkflow:a}}%
    {\phantomsubcaption\ignorespaces\label{fig:example:WashingtonWorkflow:b}}%
    {\phantomsubcaption\ignorespaces\label{fig:example:WashingtonWorkflow:c}}%
    \caption{\textbf{A typical exploration workflow with \SoftwareName:} A user can open and re-arrange views on the fly, derive new data sets using analytics plugins and connect parameters between plugins. Linked colormaps of the scatterplot and image viewer are shown in \autoref{fig:teaser}.%
    }    
    \label{fig:example:WashingtonWorkflow}%
    \vspace{-15pt}
\end{figure*}

We want to explore a hyperspectral image data set, the HYDICE image of the National Mall~\cite{WashingtonDCdata}, showing 307 by 1280 pixels, each attached to 191 spectral bands covering the \qtyrange{0.4}{2.4}{\micro\metre} region of the light spectrum reflected by the objects in view.
Each band can be interpreted as an image channel.
A major objective when exploring hyperspectral images is the identification of surface cover classes.
It is typical to manually define class labels for a small subset of pixels that afterwards are used in semi-supervised automated classification for the rest of the data.
Connecting any derived features from the spectrum back to the spatial image layout is essential during these analysis steps.
More specifically, our goals are now to (I) explore the data, connected to the task of \textit{discovering and describing observations}, and to (II) explain these observations by \textit{identifying main causes}.
These steps will yield well-justified classes that can be used in downstream analysis.

First, in \SoftwareName, we load the HYDICE data set using an \textit{image loader plugin}.
To inspect the loaded image we can open it in an \textit{image viewer plugin}, which provides single-channel and false-coloring visualizations based on any three channels.
We additionally open a \textit{spectral view plugin} which shows the full spectrum of a single pixel or the averaged spectrum of a selection that we define in the image viewer, resulting in the setup of \autoref{fig:example:WashingtonWorkflow:a}.
Then, to easily discover a hierarchical class structure, we use the \textit{HSNE analytics plugin} to create a hierarchical embedding of the data employing angular distance: we open the analysis through a context-menu of the data set entry in the \textit{data hierarchy}, select the cosine distance metric, start the embedding and display it in a \textit{scatterplot} as seen in \autoref{fig:example:WashingtonWorkflow:b}.
Next, we manually outline three clusters that are apparent in the top-level HSNE embedding as shown in \autoref{fig:teaser} (center top).
To inspect their spectra, we drag and drop the new cluster data set from the data hierarchy into the spectral viewer, \autoref{fig:teaser} (right).
Additionally, we might inspect the cluster sizes in the \textit{data properties}.
Clicking on a specific cluster displayed in the data properties will select corresponding data points in the embedding and highlight corresponding pixels in the image (\autoref{fig:example:WashingtonWorkflow:c}).
Thus, we can quickly relate the cluster spectra to image positions and define the main pixel classes water, vegetation, and buildings.
We want to focus on a single cluster \textemdash\ the one corresponding to buildings.
Therefore, we refine the cluster of interest to a lower HSNE hierarchy level through a context menu opened by clicking inside the embedding \textemdash\ the HSNE plugin added an action to the data set that is displayed there as well as in the data properties window.
To establish a visual connection between the spatial data layout and embedding, we drag the new embedding data set to the image viewer, which automatically infers the proper image dimensions for the data subset from its parent in the data hierarchy and converts it into an additional image layer.
Further, we can link the colormaps of this image layer and the embedding through the parameter-sharing system by publishing one and connecting the other to it (\ref{Req::Linkable}).
Zooming into a spatial area of interest, \autoref{fig:teaser} (left), we can discriminate between several building structures like houses and streets, and even create sub-classes of roofs that immediately stand out thanks to the embedding-based recoloring.

The above procedure intertwined the accomplishment of goals~(I) and~(II). 
\SoftwareName\ made it easy to connect various views on the data, i.e., a spatial layout, high-dimensional pixel attributes, and derived features in the form of embedding positions.
We quickly discriminated between classes in the data and identified differing spectral characteristics as their cause. 
A video that walks through the full procedure can be found as supplemental material.

\subsection{Sharing Analysis Setups -- Designer Perspective}  \label{sec:examples:applicationdesign}
\begin{figure}[b]  
    \centering
    \vspace{-2mm}%
    \pdftooltip{%
    \includegraphics[width=\linewidth]{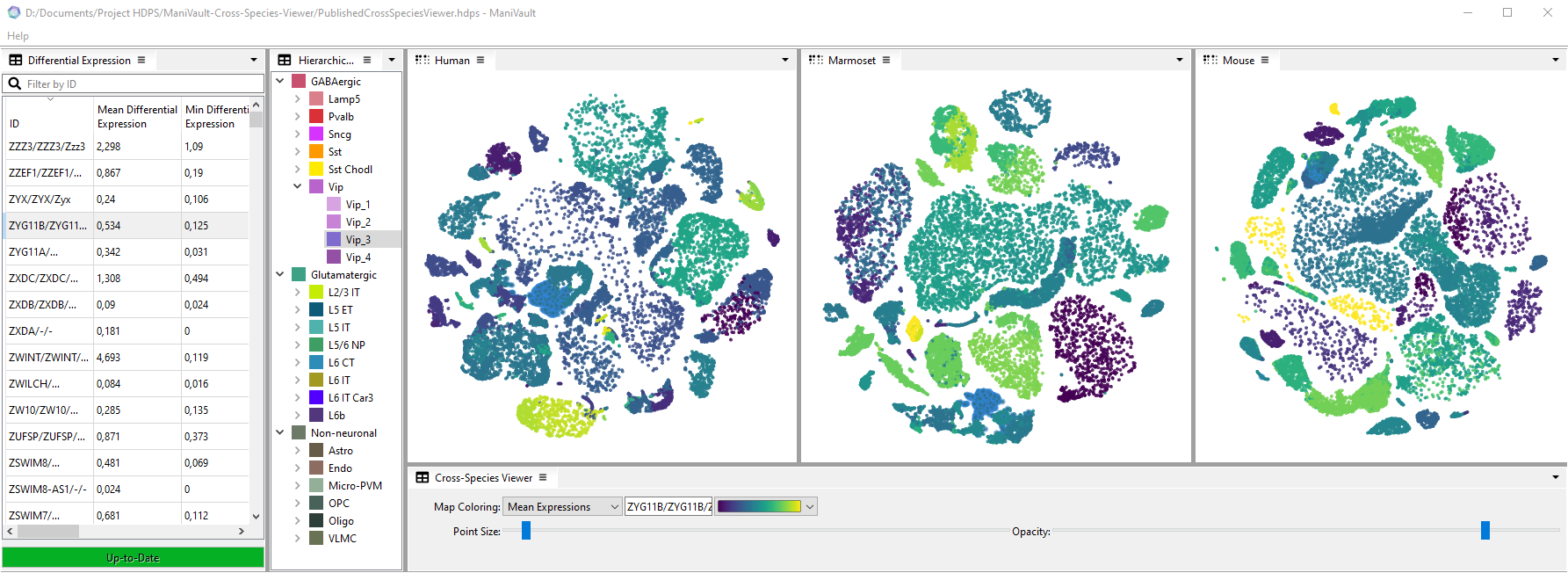}%
    }{%
    Screenshot of A ManiVault project that re-implements a Cytosplore Viewer application with three scatterplots, a table and a hierarchy view.
    }%
    \caption{Screenshot of a re-implementation of a \textbf{Cytosplore Viewer} for comparative cellular analysis of motor cortex in human, marmoset, and mouse~\cite{Bakken2021CortexMarmoset}. 
    The viewer shows embeddings of cells from the three species in combination with a shared cluster hierarchy and the option to calculate differential gene expression.
    See Suppl. S2 for a larger figure version.%
    }    
    \label{fig:example:CrossSpeciesViewer}%
\end{figure}

\SoftwareName's workspace and project features can be used to save and continue an analysis session but also enable dissemination of results and complete workflows.
To showcase this, we re-implemented the Cytosplore Viewer application~\cite{CytosploreViewer} dedicated to sharing the results of {Bakken et al.}~\cite{Bakken2021CortexMarmoset} in \SoftwareName, shown in \autoref{fig:example:CrossSpeciesViewer}.
Instead of having to write an entire stand-alone application to share an interactive environment alongside data to explore related insights in, we can use a \SoftwareName\ project to bundle both views and data (\ref{Req::Distributable}).

The viewer application depicts RNA sequencing data on brain cells from three vertebrate species.
The viewer aims to highlight differences in the expression of genes and cell types that are shared across the species as described in the original paper.
The main elements of the viewer application are three scatterplots showing t-SNE embeddings of the gene data of each species, a hierarchical cluster viewer showing cell types, and a table view showing statistical properties of the expression data.
To create the viewer, we configure \SoftwareName's GUI from within the GUI (\ref{Req::Configurable}).
We start with loading all data sets and setting up a single scatterplot plugin.
We link scatterplot parameters like its colormap to a global settings panel that lets users configure all three scatterplots, like in the original application.
Its settings can be saved as a preset which we use for the other two scatterplot instances.
Similarly, we populate the cluster hierarchy view and table viewer with data.
\Cref{fig:example:CrossSpeciesViewer} shows a configuration in which a user-selected entry in the table view defines the data attributes (here a gene's expression) used to recolor the scatterplot data points (here tissue samples).

\SoftwareName's Studio Mode allows us to lock this setup of views and parameter connections. 
This is achieved by simply publishing the current view layout, loaded data, and parameter linkage through the "File" menu tab.
We can now share the viewer with other parties.

\section{Discussion and Conclusion} \label{sec:conclusion}

This paper describes the design considerations for and implementation of \SoftwareName, an extensible visual analytics framework for high-dimensional data.
Due to its modular architecture and data-centric design, the software enables flexible exploration and analysis workflows.
We presented various plugins that provide visualization and analytics functionalities to the system.
To build upon these, we showed how existing libraries can be easily incorporated into the system.
\SoftwareName's action and event systems allow users to adjust plugins and their interplay, enabling the creation of fully customized applications.

Currently, the system provides data plugins that cover a wide range of applications.
New data types like multivariate graph data~\cite{Kerren2014MultivariateNetwork} can be introduced into the system as new data plugins without changes to the application's core.
We plan to extend the current serialization mechanism, used for saving the state of the system, to handle information about interaction history and other kinds of provenance~\cite{Ragan2016Provenance}.
Finally, we would like to include analytics plugins that run code in interpreted languages like Python or R, to easily integrate the vast amount of data science tools available in those languages.

We believe that \SoftwareName\ has great potential in aiding with the creation and use of visual analytics applications for visualization developers, practitioners, and application designers.

\checkMaxPages{9} %

\section*{Supplemental Materials}
\label{sec:supplemental_materials}
All supplemental materials are available on OSF at \url{https://osf.io/9k6jw/}, released under a \href{https://creativecommons.org/licenses/by-sa/4.0/}{CC BY SA 4.0} license.
In particular, they include (1) benchmark results, S1, and a larger version of \autoref{fig:example:CrossSpeciesViewer}, S2, (2) Excel files containing the data presented in S1, (3) Python scripts to run the nptsne benchmark from S1, (4) two videos showcasing ManiVault and (5) a full version of this paper.

\acknowledgments{%
Author contributions: Alexander Vieth: Writing and Plugin Development; %
Thomas Kroes: Lead Developer (Core and Plugins) \& Architect; %
Julian Thijssen: Developer \& Architect of initial core, Plugin Developer; %
Baldur van Lew: Build Infrastructure; %
Jeroen Eggermont and Soumyadeep Basu: Viewer \& Plugin Development; %
Elmar Eisemann, Anna Vilanova, Thomas Höllt, and Boudewijn Lelieveldt: project conception, manuscript writing, general supervision.

This work received financial support from %
the NWO TTW project 3DOMICS (NWO: 17126), %
the NWO Gravitation project \mbox{BRAINSCAPES}: A Roadmap from Neurogenetics to Neurobiology (NWO: 024.004.012), and %
the NIH Brain Initiative Cell Atlas Network (UM1MH130981).
}

\bibliographystyle{abbrv-doi-hyperref}

\bibliography{references}

\checkMaxPages{11} %

\clearpage

\setcounter{page}{1}
\onecolumn
\begin{center}
{\large Supplemental Material: \\}
\vspace{5pt}

\makeatletter
{\sffamily\iftoggle{vgtc@journal}{\huge}{\LARGE\bfseries}%
        \vgtc@sectionfont%
        \vgtc@title \par}
\makeatother

\vspace{14pt}
\makeatletter
{\large\sffamily\vgtc@sectionfont
        \vgtc@author}

\makeatother

\vspace{14pt}
\end{center}

\renewcommand{\figurename}{Supplementary Fig.}
\renewcommand{\tablename}{Supplementary Table}

\section*{S1: Benchmarks} \labelshort[S1]{supp:benchmarks}

\paragraph{Speed}
\SoftwareName\ can show progressive updates of analytics plugins with only small additional computational penalties. 
To show this, we compute t-SNE embeddings with a \SoftwareName\ analytics plugin that uses the HDI library GPGPU implementation of t-SNE~\citesupplement{supp:pezzotihdi, supp:Pezzotti2019Gpgpu}.
First, we compute embeddings non-progressively, and then, in a second setting, we show intermediate embeddings every 10 gradient descent iterations (respectively "no updated" and "with updates" in \autoref{supp:table:timings}.
Additionally, we compare these runs with a lightweight python wrapper~\citesupplement{supp:Lew2021Nptsne} around the same t-SNE library.
Every embedding is laid out over 500 gradient descent iterations.
The non-progressive computation is slightly faster than the Python wrapper around the same library calls.
The difference between the total runtime of the t-SNE embeddings in \SoftwareName\ with and without updates is explained by the difference in the gradient descent time: In the former setting, the analytics plugin notifies \SoftwareName's core about the current embedding layout.
All measurements were taken on a machine equipped with an NVIDIA GeForce RTX 2080 SUPER GPU and an Intel Core i5-9600K CPU and running Windows 11 22H2.
The supplemental material "benchmark\_time.xlxs" contains the full data.

\begin{table}[h]
\centering
\begin{threeparttable}
\caption{Duration of t-SNE embedding computations with the same implementation, invoked via a Python wrapper and \SoftwareName, once showing only the final embedding and once progressively updating a scatterplot. 
Times, in seconds, are averages over 10 runs with sample standard deviation.}
\label{supp:table:timings}
\begin{tabular}{l c c c c c}
\toprule
\multicolumn{1}{r}{Data set} & 
Swiss Roll 3D~\citesupplement{supp:Tenenbaum1997} & 
COIL-20~\citesupplement{supp:Nene1996} & 
MNIST~\citesupplement{supp:LeCun1998} &
Fashion-MNIST~\citesupplement{supp:xiao2017} & 
10x Mouse~\citesupplement{supp:Zheng2017} \\
\multicolumn{1}{r}{\footnotesize\# points} & 
{\footnotesize1,500} & %
{\footnotesize1,440} & %
{\footnotesize70,000} & %
{\footnotesize70,000} & %
{\footnotesize1,306,127} \\ %
\multicolumn{1}{r}{\footnotesize\# dimensions} & 
{\footnotesize3} & %
{\footnotesize16,384} & %
{\footnotesize784} & %
{\footnotesize784} & %
{\footnotesize50 (first PCs)} \\ %
\midrule
nptsne~\citesupplement{supp:Lew2021Nptsne} (Python wrapper) &
0.30 \, {\footnotesize(0.02)} &  %
2.32 \, {\footnotesize(0.08)}  & %
23.31 \, {\footnotesize(0.14)} & %
20.58 \, {\footnotesize(0.01)} & %
268.38 \, {\footnotesize(2.21)} \\ %
ManiVault (no updates)  & 
0.58 \, {\footnotesize(0.05)} &  %
2.37 \, {\footnotesize(0.05)} &  %
22.51 \, {\footnotesize(0.11)} & %
20.20 \, {\footnotesize(0.27)} & %
258.60 \, {\footnotesize(5.76)} \\ %
ManiVault (with updates) &
0.59 \, {\footnotesize(0.07)} &  %
2.46 \, {\footnotesize(0.09)} &  %
22.85 \, {\footnotesize(0.15)} & %
20.24 \, {\footnotesize(0.11)} & %
257.91 \, {\footnotesize(4.02)} \\ %
\bottomrule
\end{tabular}
\begin{tablenotes}[para]\footnotesize
  Note: Times in seconds, sample standard deviation in parentheses.
\end{tablenotes}
\end{threeparttable}
\end{table}

\paragraph{Memory}
After starting \SoftwareName, with the Data Hierarchy and Data Property Viewer open, the software consumes around $87$ MB of memory (on Windows).
Loading data sets comes with a small memory overhead.
Here, we loaded various data sets, as listed in \autoref{supp:table:timings}, and compared their binary size on disk with the growing memory footprint of \SoftwareName\ after loading them.
We observe a $0.7-1.5$ MB overhead per data set, compared to their binary size, when utilizing the point data type plugin.
For larger data set, it can be useful to trade off precision for lower memory uptake. 
We can employ a \href{https://en.wikipedia.org/wiki/Bfloat16_floating-point_format}{bfloat16} floating point implementation~\citesupplement{supp:Dekker2020bfloat} to store large data set, and thereby effectively half the memory \SoftwareName\ requires: e.g. the 10x Mouse data will take up $126.15$ MB instead of $249.87$ MB.
The supplemental material "benchmark\_memory.xlxs" contains the full data.

\begin{table}[h]
\centering
\begin{threeparttable}
\caption{Memory consumption of loaded data sets, as listed in \autoref{supp:table:timings}, in ManiVault compared to their binary size on disk. 
Values are averages over 4 loaded data sets.}
\label{supp:table:memory}
\begin{tabular}{l c c c c c}
\toprule
\multicolumn{1}{r}{Data set} & 
\citesupplement{supp:Tenenbaum1997} & 
\citesupplement{supp:Nene1996} & 
\citesupplement{supp:LeCun1998} &
\citesupplement{supp:xiao2017} &
\citesupplement{supp:Zheng2017} \\
\multicolumn{1}{r}{\footnotesize Binary type} & 
{\footnotesize float32} & %
{\footnotesize uint8} & %
{\footnotesize uint8} & %
{\footnotesize uint8} & %
{\footnotesize float32} \\ %
\midrule
Raw binary &
0.017 &  %
22.5  & %
52.34 & %
52.34 & %
249.12 \\ %
ManiVault (float32) & 
0.97 &  %
- &  %
- & %
- & %
249.87 \\ %
ManiVault (uint8) &
- &  %
23.77 &  %
53.5 & %
54.1 & %
- \\ %
\bottomrule
\end{tabular}
\begin{tablenotes}[para]\footnotesize
  Note: Values in MB. Slight deviations might occur due to Qt's memory management on Windows 11, e.g., the 
  difference between MNIST and Fashion-MNIST. 
\end{tablenotes}
\end{threeparttable}
\end{table}

\newpage
\section*{S2: Larger Figures} \labelshort[S2]{supp:largeFigures}

\begin{figure*}[h]  
    \centering
    \pdftooltip{%
    \includegraphics[width=\linewidth]{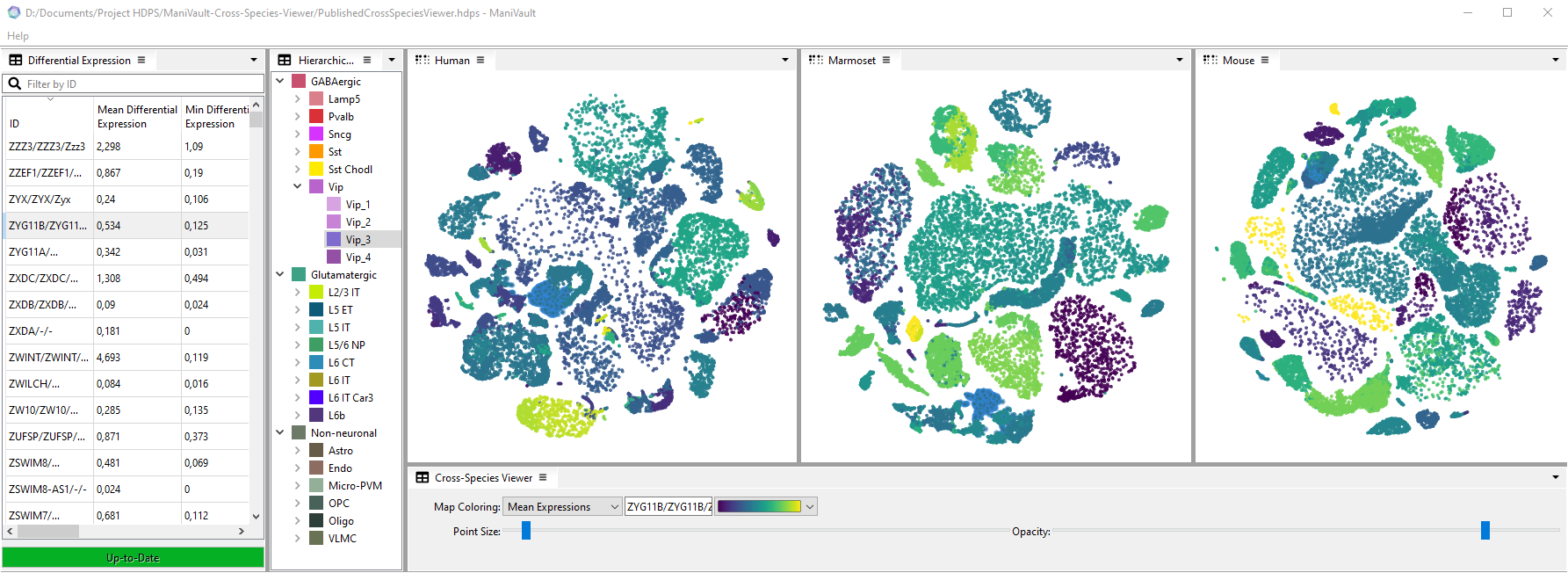}%
    }{%
    Screenshot of A ManiVault project that re-implements a Cytosplore Viewer application with three scatterplots, a table and a hierarchy view.
    }%
    \caption{Large version of \autoref{fig:example:CrossSpeciesViewer}.%
    }    
    \label{fig:supp:CrossSpeciesViewer}%
\end{figure*}

\bibliographystylesupplement{abbrv-doi-hyperref-narrow}
\bibliographysupplement{references}

\end{document}